\documentclass[final,3p]{elsarticle}

\usepackage{amsmath}
\usepackage{bm}
\usepackage{amsfonts}
\usepackage{mathtools}
\usepackage{hyperref}
\usepackage{subcaption}
\usepackage{tikz}
\usepackage{algorithm}
\usepackage{algpseudocode}
\usepackage{lineno}
\usepackage{ulem}
\linenumbers

\DeclareMathOperator{\conv}{conv}
\newcommand{\rom}[1]{{\scshape{#1}}}

\newcommand{\reviewerone}[1]{{\color{black}#1}}
\newcommand{\NewAnswers}[1]{{\color{black}#1}}
\newcommand{\reviewertwo}[1]{{\color{black}#1}}

\journal{arXiv}

\begin{document}
\nolinenumbers

\begin{frontmatter}

\title{Parallel athermal quasistatic deformation stepping of molecular systems}

\author[uniSTG]{Maximilian Reihn}
\affiliation[uniSTG]{organization={Institute for Applied Analysis and Numerical Simulation, University Stuttgart},
            addressline={Pfaffenwaldring 57}, 
            city={Stuttgart},
            postcode={70569}, 
            state={Baden-Württemberg},
            country={Germany}}

\author[rwth]{Franz Bamer}%
\affiliation[rwth]{organization={Institut für Allgemeine Mechanik},
            addressline={Eilfschornsteinstraße 16}, 
            city={Aachen},
            postcode={52062}, 
            state={Nordrhein-Westfalen},
            country={Germany}}
            
\author[uniSTG]{Benjamin Stamm}

\begin{abstract}
The athermal quasistatic deformation method provides an elegant solution to overcome the limitation of short time spans in molecular simulations. It provides overdamped conditions, allowing for the extraction of purely structural responses in the absence of thermal vibration.
However, it requires computationally expensive sequences of affine deformation followed by minimization of the potential energy to incrementally find the path in the potential energy landscape that corresponds to the correct solution trajectory.
Therefore, we propose an athermal parallel stepping scheme that significantly improves the computational time necessary to find the correct solution trajectory using a multi-thread approach.
Our approach proposes stepping at two levels. Level \rom{I} stepping provides a sequence of initial guesses at large increments by affine deformation of the system and land-marking anchor points on the potential energy landscape. Level \rom{II} stepping performs a set of individual finely resolved athermal quasistatic deformation steps between the inherent structures of the initial level \rom{I} guesses executed in parallel. The evaluated candidate trajectory is then verified by consecutively comparing the configuration of every last level \rom{II} result with the corresponding inherent structure of the level \rom{I} guesses at the same strain states. If the two configurations are not equivalent, the solution must be rejected and recalculated from this point.
Rigorous numerical testing with $4,8,16$ and $32$ parallel threads and \reviewerone{different values of hyper-parameters} demonstrates that our method achieves computational \reviewerone{average} speed-ups of factors ranging from \reviewerone{$2.02$ to $6.33$}, while maintaining simulation accuracy, offering a powerful new tool for athermal molecular simulations.
\end{abstract}

\begin{graphicalabstract}
\includegraphics{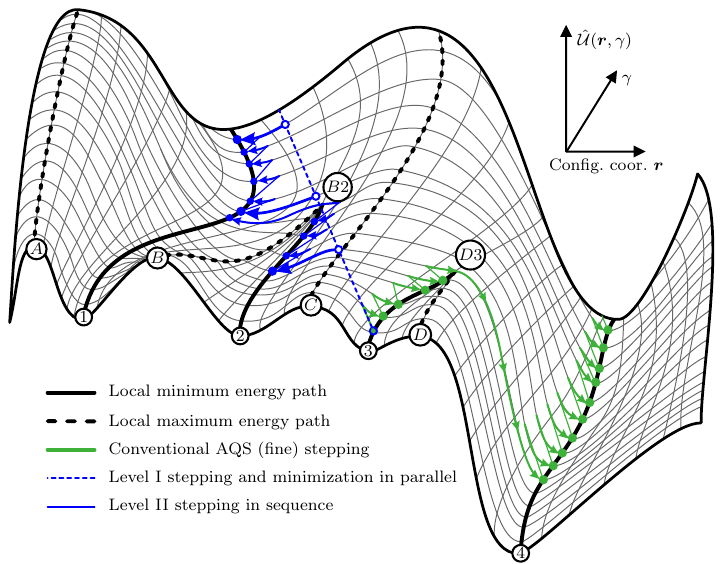}
\end{graphicalabstract}

\begin{highlights}
\item We propose a novel two-level athermal parallel stepping scheme that accelerates quasistatic molecular simulations.
\item With the help of multi-threading and characteristics of the potential energy surface of molecular systems, we perform athermal deformation protocols at different deformation states simultaneously.
\item We achieve consistent computational speed-ups, with an average acceleration factor of between \reviewerone{$2.02$ and $6.33$, depending on the number of threads $P$ and parameter $K$} used, compared to conventional athermal quasistatic deformation methods, while preserving the same level of accuracy.
\end{highlights}


\end{frontmatter}
 


\section{Introduction}
\label{sec:introduction}
Molecular models provide valuable insights into the structure-property relationship of nanostructures and give us a general idea of how materials respond to external deformation \cite{Rappaort2004}.
In particular, mechanical phenomena, such as fracture, originate from the nanoscale and span multiple orders of magnitude, connecting nanoscale inelastic behavior to larger-scale dynamics \cite{DAL2022115639}.

Molecular models are built on the assumption that atoms are point masses, while appropriate potential functions describe their interactions. They allow for significantly larger models than quantum-mechanical frameworks and often serve as a bridge to continuum-mechanical descriptions \cite{raabe1998}. However, physically meaningful studies require a large number of atoms, so the dimension of the problem is high, and the number of particle interactions to be considered is even higher. These issues limit the application of molecular models and make an overall efficient implementation challenging \cite{Nos1984AUF, Andersen1980MolecularDS}.

Significant effort has been devoted to overcoming this issue, with proposals that can be categorized into two groups. The first group includes methods that optimize the distance evaluations and interaction evaluation by introducing cutoff radii, Verlet lists, and Linked Cell lists \cite{Rappaort2004, DOBSON2016211}. The second group of methods focuses on multi-threading, i.e., distributing the computation effort within several tasks that are executed individually \cite{plimpton55, TRANCHIDA2018406}. Such parallelization can be performed through atom-wise, force, and spatial decomposition, whereas the third strategy has shown the highest efficiency for molecular problems \cite{Plimpton1993FastPA, Phillips2005ScalableMD}.

Notably, such strategies accelerate computation during a single incremental calculation step, typically performed during a time integration algorithm or a procedure that minimizes the potential energy. Due to the high dimensionality of molecular models, explicit time integration schemes, such as the Velocity Verlet algorithm \cite{Verlet1967, Grubmller1991GeneralizedVA}, have been established as the preferred algorithms since they neither require iterations during every step nor the assembly and subsequent inversion of the Hessian matrix. An essential disadvantage of these methods is, however, that the computation time step chosen must not be larger than the highest eigenfrequency of the linearized system at the current deformation state \cite{FINCHAM1986263, Ormeno2024}. Effectively, this value is proportional to the vibration period of the strongest atomic bonds, forcing one to choose calculation time steps in the range of femtoseconds. Consequently, a disproportionately large number of calculation steps are required to cover a given simulation time period, severely limiting the possible simulation time span. Therefore, molecular simulations are not only limited by system size but also by the simulation time window, greatly hindering up-scaling.

When performing mechanical deformation, the athermal quasistatic simulation or zero-temperature method \cite{Maloney2004} alleviates the limiting time problem to some extent, as it allows for infinitely slow mechanical deformation in the absence of Brownian motion. It describes an overdamped system, continuously following the minimum configuration of the atomic system. However, it requires subsequent minimization procedures during every deformation step, leading again to a limitation due to computational effort.
Thus, this paper proposes a parallel athermal quasistatic stepping algorithm using multi-thread machines. Notably, our method provides exact simulation results, saving computational time compared to conventional athermal quasistatic deformation stepping while being flexible enough so that the force evaluation can be further parallelized on top.

The paper is organized as follows. First, the athermal quasistatic deformation method is introduced, along with a benchmark example demonstrating the parallel deformation stepping method. Subsequently, we present a new parallel computing approach, supported by rigorous studies that demonstrate computational benefits over the conventional athermal quasistatic deformation method. Finally, conclusions are drawn, and an outlook for future work is provided.

\section{\label{sec:AQS}Athermal quasistatic deformation}
Being exclusively interested in the structural response of the material, we investigate mechanical deformation without considering any thermal vibrations \cite{HUFNAGEL2016375,moldynKrishna,raabe1998}, so the configuration remains in a local minimum of the potential energy landscape \reviewerone{sometimes also known as the potential energy surface (PES)}. Effectively, we are investigating materials at zero temperature. \reviewerone{This type of simulation is called athermal quasistatic (AQS) deformation protocol.}  Such an assumption remains meaningful as long as we consider materials at temperatures significantly lower than the melting temperature or, in the particular case of glasses, considerably lower than the glass transition temperature \cite{Falk1998}.
\reviewerone{\subsection{Formalism}}
We confine a total number of $N$ atomic particles into a $d$-dimensional simulation cell whose geometry is defined by the second-order Bravais tensor 
\reviewerone{$\hat{\bm{H}}=\left[\hat{\bm{h}}_1,\hat{\bm{h}}_2\right]\in \mathbb R^{2\times 2}$ for dimension} $d=2$ and \reviewerone{$\hat{\bm{H}}=\left[\hat{\bm{h}}_1,\hat{\bm{h}}_2,\hat{\bm{h}}_3\right]\in \mathbb R^{3\times 3}$} for $d=3$, where $\hat{\bm{h}}_1,\hat{\bm{h}}_2$ and $\hat{\bm{h}}_3$ refer to the linearly independent Bravais vectors \cite{Thompson2009GeneralFO}. \reviewerone{In the following manuscript, the cell shape inducing Bravais tensor is also referred to as a matrix. With the linear independence of the Bravais vectors, the matrix inverse of $\hat{\bm{H}} \in \mathbb R^{d\times d}$ always exists as $\left(\hat{\bm{H}}\right)^{-1} \in \mathbb R^{d\times d}$.} Periodic boundary conditions are applied in the direction of all three Bravais vectors, although the method proposed in this paper also applies to systems without periodic boundary conditions \cite{BARCLAY2021110238}.

External deformation is performed by altering the shape of the simulation cell through the manipulation of Bravais vectors, assembly of the Bravais tensor, and adjustment of atomic positions \reviewertwo{and all respective boundary conditions}. We define a configuration \reviewerone{$\bm H^{(\gamma)}\in \mathbb R^{d\times d}$}, which was deformed by an affine deformation parameter $\gamma\in\mathbb{R}$ from the reference configuration \reviewerone{$\hat{\bm{H}}\in \mathbb R^{d\times d}$}. The deformation is done incrementally by multiplying the deformation tensor $\bm{F}^{(\Delta\gamma)}$ \reviewertwo{acting on $\bm{H}^{(\gamma)}$} such that $\bm{H}^{(\gamma+\Delta\gamma)}=\bm{F}^{(\Delta \gamma)} \bm{H}^{(\gamma)}$. For $\gamma=0$ the \reviewerone{the matrix inducing the } current configuration is \reviewerone{equal to the matrix inducing} the reference configuration, meaning \reviewertwo{$\bm{H}^{(\gamma=0)}=\hat{\bm H}$}. Mechanical deformation is induced by altering the Bravais tensors, written as:
\begin{align}
    \bm{F}^{(\Delta \gamma)} = \bm{H}^{(\gamma+\Delta\gamma)} \left(\bm{H}^{(\gamma)}\right)^{-1} \;.
\end{align}
Atoms are defined by their position vectors within the configuration
$${\bm{r}}_i \in \conv(\bm{H}) \coloneqq
\conv\left( \left\{ \, \sum_{j=1}^d\alpha_j \mathbf{h}_{j} \;\middle|\; \forall (\alpha_1,\ldots,\alpha_d)\in\{0,1\}^d \right\}\right)
=\conv\left( \left\{ \, \bm H\bm \alpha \;\middle|\; \forall \bm\alpha\in\{0,1\}^d \right\}\right)
$$
for all $i=1,\ldots,N$ where we stack all the atomic positions to one concatenated position vector $\bm{r} \in \conv(\bm{H})^N\coloneqq \conv(\bm{H}) \times \ldots \times \conv(\bm{H}) \subset  \mathbb{R}^{dN}$. The expression \reviewerone{\reviewertwo{$\conv$ of a set of points describes the (standard) convex hull of those points.}} 

The mapping $\bm{F}$ \reviewerone{for a shear step size $\Delta\gamma$} is applied to a single particle $i$'s position as follows:
\begin{align}
    \begin{split}
        \bm{F}^{(\Delta \gamma)}: \conv(\bm{H}^{(\gamma)}) &\rightarrow \conv(\bm{H}^{(\gamma+\Delta\gamma)}),\\
        \bm{r}_i&\mapsto \bm{F}^{(\Delta \gamma)} \bm{r}_i\;.
    \end{split}
\end{align}
The deformation mapping for the entire $N$-particle system is written as:
\begin{align}\label{eq:Npower}
    \begin{split}
         \bm{F}^{(\Delta \gamma)}_N: \conv(\bm{H}^{(\gamma)})^N &\rightarrow \conv(\bm{H}^{(\gamma+\Delta\gamma)})^N,\\
        \bm{r}&\mapsto 
        \bm{F}^{(\Delta \gamma)}_N \bm{r}
        :=
        \begin{bmatrix}
            \bm{F}^{(\Delta \gamma)} & & \bm 0 \\
             & \ddots &  \\
            \bm 0&  & \bm{F}^{(\Delta \gamma)}
        \end{bmatrix}
        \reviewerone{
        \begin{bmatrix}
            \bm{r}_1 \\
             \vdots \\
            \bm{r}_N
        \end{bmatrix}.}
    \end{split}
\end{align}
In the following, we observe the system from the Lagrangian perspective \cite{AN2021110338}, which means that we formulate our equations with variables associated with the reference configuration $\bm{\hat{H}}$. Note that $\bm{F}^{(\gamma)}$ provides the mapping from the reference $\hat{\bm{H}}$ to the current configuration $\bm{H}^{(\gamma)}$.
We first define the potential energy landscape $\mathcal{U}$ as 
\begin{align}
    \begin{split}
        \mathcal{U}(\bm{r}):\reviewerone{\conv(\bm{H}^{(\gamma)})^N} &\rightarrow \mathbb{R}\\
        \bm{r}&\mapsto \mathcal{U}(\bm{r}),
    \end{split}
\end{align}
depending on the atomic positions $\bm{r}$. In order to keep track of the deformation, we also define the potential energy landscape with respect to the reference configuration as 
\begin{align}
    \begin{split}
        \hat{\mathcal{U}}(\hat{\bm{r}},\gamma)\coloneqq \mathcal{U}(\bm{F}^{(\gamma)}\bm{\hat{r}}):\conv(\hat{\bm{H}})^{N}\times \mathbb{R} &\rightarrow \mathbb{R}\\ \hat{\bm{r}}&\mapsto \hat{\mathcal{U}}(\hat{\bm{r}},\gamma) \; .
    \end{split}
\end{align}
Since the configuration remains in a minimum of the potential energy landscape \cite{Sastry2000TheRB, Heuer2008ExploringTP,Payne1992IterativeMT}, the force $\hat{f}\in \mathbb{R}^{dN}$, which is equal to the negative gradient \reviewerone{of the potential energy} with respect to position, vanishes:
\begin{align}
    \hat{f}(\hat{\bm{r}},\gamma) \coloneqq -\nabla_{\hat{\bm{r}}} \,  \hat{\mathcal{U}}(\hat{\bm{r}},\gamma) = 0 \; , \quad \text{with} \quad \nabla_{\hat{\bm{r}}} \coloneqq
    \left(
    \begin{array}{c}
        \frac{\partial }{\partial \hat{\bm{r}}_1} \\
        \vdots \\
        \frac{\partial }{\partial \hat{\bm{r}}_{dN}}
    \end{array}
    \right) \; .
\end{align}
Mechanical deformation is induced by altering the Bravais vectors from the reference configuration $\hat{\bm{H}}$ to the current configuration $\bm{H}^{(\gamma)}$. Therefore, also altering the potential energy landscape. However, to ensure mechanical equilibrium, the configuration must remain in a local minimum. Considering that $\hat{f}$ depends on $\bm{H}^{(\gamma)}$, and $\bm{r}$ itself depends on $\bm{H}^{(\gamma)}$, we evaluate the change in force from the reference to the current configuration as the total derivative:

\reviewertwo{
\begin{align}
    \label{eq:equilibrium}
    \frac{d \hat{f}(\hat{\bm{r}},\gamma)}{d\gamma} = -\frac{\partial }{\partial \gamma}\nabla_{\hat{\bm{r}}}\hat{\mathcal{U}}\left(\hat{\bm{r}},\gamma\right) - \bm{\mathcal{\bm{\hat{H}}}}\left(\hat{\bm{r}},\gamma\right) \frac{d\hat{\bm{r}}}{d\gamma} \; ,
\end{align}
}
where $\bm{\mathcal{\hat{\bm{H}}}}\left(\hat{\bm{r}},\gamma\right) =\frac{\partial^2\mathcal{\hat{U}}\left(\hat{\bm{r}},\gamma\right)}{\partial \hat{\bm{r}} \partial \hat{\bm{r}}}$ refers to the Hessian of the potential energy landscape, depending on both the positions and the \reviewerone{current deformation according to the deformation parameter $\gamma$}. The change in potential energy with respect to the cell geometry may be interpreted as a virtual force \reviewertwo{$\Tilde{F}=-\frac{\partial }{\partial \gamma}\nabla_{\hat{\bm{r}}}\hat{\mathcal{U}}\left(\hat{\bm{r}},\gamma\right)$} pushing the externally disturbed configuration back into the minimum position. The term $\frac{d\hat{\bm{r}}}{d\gamma}$ refers to a change in the reference position due to the external \reviewerone{deformation}, which \reviewerone{induces a displacement field} also referred to as non-affine deformation field \cite{Maloney2005AmorphousSI,Zaccone2013ELASTICDI,Zaccone2012DisorderassistedMA}.

During mechanical deformation, we demand that a change in the force with respect to mechanical deformation is also equal to zero, $\frac{d \hat{f}(\hat{\bm{r}},\gamma)}{d\gamma} = 0$, leading to the equation of equilibrium for molecular systems:
\begin{align}
    \Tilde{F} = \bm{\mathcal{\hat{H}}}\left(\hat{\bm{r}},\gamma \right) \frac{d\hat{\bm{r}}}{d\gamma} \; .
\end{align}
\reviewerone{This} means that an infinitesimally disturbed system experiences a non-affine displacement field $\frac{d\hat{r}}{d\gamma}$ so that a force $\Tilde{f}$ is required to push the system back into its local minimum.

\begin{figure}[htbp]
    \centering
    \includegraphics[width=0.8\linewidth]{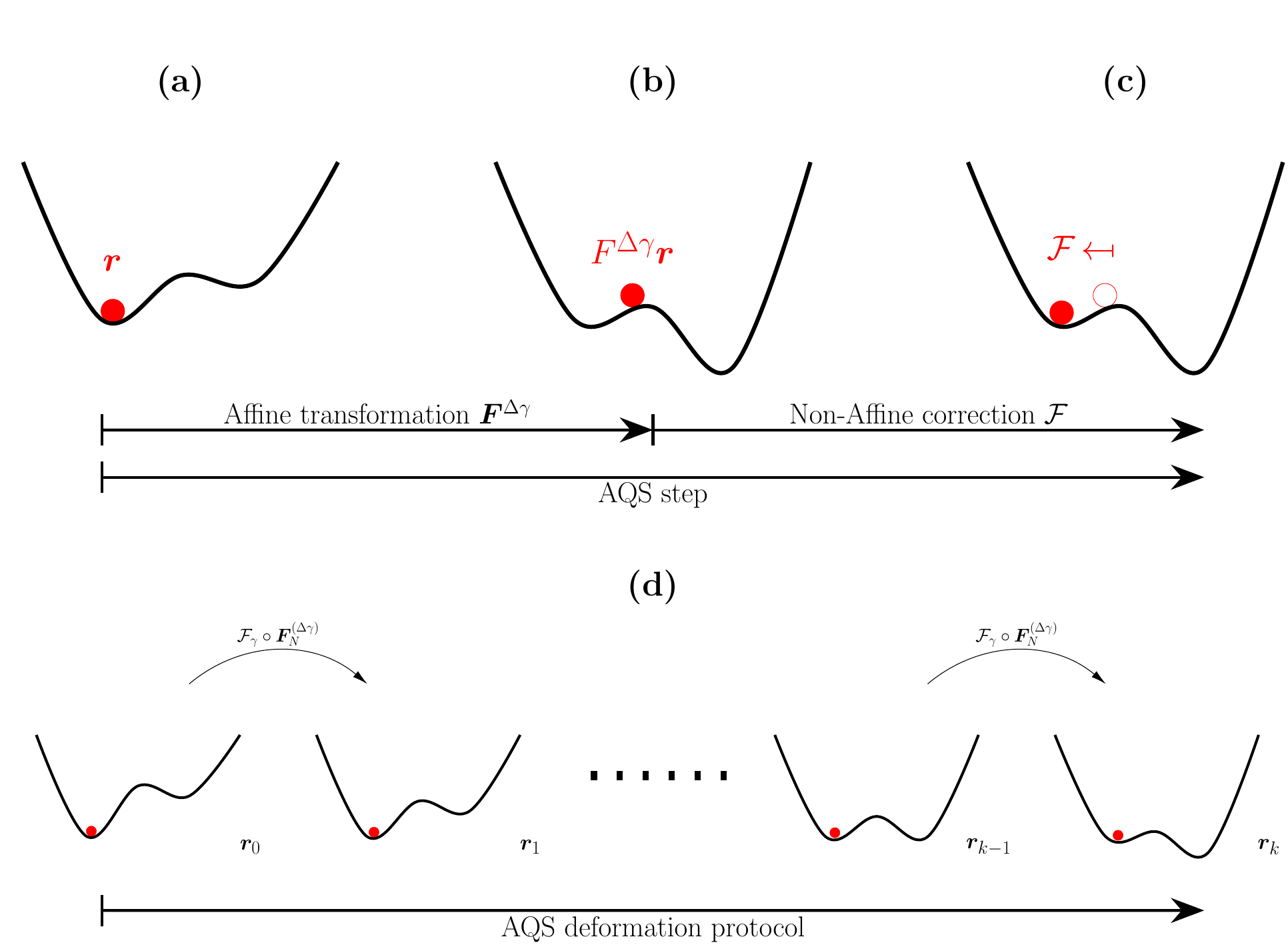}
    \caption{\reviewerone{Illustration of the athermal quasistatic deformation on a one-dimensional potential energy landscape; (a) undeformed configuration; (b) configuration after affine displacement mapping; (c) non-affine correction; (d) and entire athermal deformation protocol of $k$ steps.}}
    \label{fig:aqs_protocol}
\end{figure}

The numerical implementation of the athermal deformation scheme is a finite \reviewerone{discretization} of Equation (\ref{eq:equilibrium}) performed within a
two-step process. Figure \ref{fig:aqs_protocol}(a,b,c) visualizes the potential energy landscape during one deformation step. Figure \ref{fig:aqs_protocol}(a) presents the potential energy landscape belonging to the reference simulation cell. The red ball represents the reference position.
In the first step, the configuration \reviewerone{$\bm{H}^{(\gamma=0)}$} is externally deformed, inducing small affine increments to $\bm{H}^{(\Delta\gamma)}$ and $\bm{r}$ using $\bm{F}^{(\Delta\gamma)}$ \reviewerone{and $\bm{F}^{(\Delta \gamma)}_N$ respectively}, altering the shape of the potential energy landscape. Generally, this leads to the system being pushed out of the local minimum so that we are not in mechanical equilibrium anymore, as indicated in Figure \ref{fig:aqs_protocol}(b). In the second step, the shape of the deformed simulation cell $\bm{H}^{(\Delta\gamma)}$ is held, while the configuration $\bm{r}$ is allowed to relax so that it again falls back into the adjacent local minimum of the potential energy landscape. This relaxation refers to the non-affine portion of the structural response and is realized by minimizing the potential energy landscape. \reviewerone{To formalize the minimization we first define the basin of attraction of $\bm{r}_{\infty}$ as the set of all configurations $\bm{r}$ which will converge to the same limit 
\begin{align}\label{eq:basin_of_attr} 
    \begin{split}
        \Omega_{\bm r_{\infty},\gamma}\coloneqq\left\{ \bm{r} \in \mathbb \conv(\bm{H}^{(\gamma)})^N\;\middle\vert\; \bm{r}_{\infty}=\lim_{t\rightarrow \infty} \bm{r}(t) \,\,\text{ with } \tfrac{d}{dt} \bm{r}(t) = -\nabla_{\bm{r}} \,\,\mathcal{U}(\bm{r}(t)) ;\,\,\bm{r}(0)=\bm{r}  \right\}.
    \end{split}
\end{align}
} The minimization \reviewerone{for a configuration $\bm{r}\in\Omega_{\bm{r}_{\infty},\gamma}$} is then defined as:
\begin{align}\label{eq:mini_procedure} 
    \begin{split}
        \mathcal{F}:\conv(\bm{H}^{(\gamma)})^{N} &\rightarrow \conv(\bm{H}^{(\gamma)})^{N},\\
        \bm{r} &\mapsto \bm{r}_{\infty},
    \end{split}
\end{align}
which is the limit of the gradient flow\reviewerone{, i.e., the continuous limit of the gradient descent, }and ends up in the adjacent local minima. \reviewerone{Note, that the minimization $\mathcal{F}$ depends on the shear value of $\gamma$, however, this is implicitly given by the space in which the minimization is performed (the same argument can be made about the evaluation of $\mathcal{U}$).}
 Generally we use a conjugate gradient method with $\bm{r}$ as the initial guess.
\\

For ease of notation, we also use the iterated concatenation, i.e. 
\begin{align}
    \left(\mathcal{F} \circ \bm{F}^{(\Delta \gamma)}_N\right)^n (\bm{r}) \;\coloneqq \; \mathcal{F}\left(\bm{F}^{(\Delta \gamma)}_N\left(\mathcal{F}\left(\bm{F}^{(\Delta \gamma)}_N\left( \ldots \bm{F}^{(\Delta \gamma)}_N \left( \bm{r} \right) \ldots  \right)\right)\right)\right)\;.
\end{align}
Using that notation \reviewerone{ we have established}, we combine the two steps into a single AQS step $\mathcal{F}\circ \bm{F}^{(\Delta \gamma)}_N$. Repeating \reviewerone{this operator} in a sequence results in \reviewerone{the formal definition of the} athermal quasistatic deformation protocol:
\begin{align} \label{eq:AQS_iteration}
    \bm{r}_k \coloneqq (\mathcal{F}\circ \bm{F}^{(\Delta \gamma)}_N)(\bm{r}_{k-1})= (\mathcal{F}\circ \bm{F}^{(\Delta \gamma)}_N)^k (\bm{r}_0) \; ,
\end{align}
where $\bm{r}_k$ is the particle configuration in the respective local minimum of the potential energy landscape after $k$ deformation steps which is confined in the space $\conv(\bm{H}^{(\gamma)})^N$ for $\gamma = k\Delta\gamma$. Per construction the deformation happens in equidistant steps of $\Delta\gamma$.

To calculate the configuration at deformation step $k$, $\bm{r}_k$, one must follow the sequence of $k-1$ times the affine mapping followed by the non-affine correction, as shown in Figure \ref{fig:aqs_protocol}(d). Also, here, the configuration is indicated by the red ball. It runs along a path of local minima in the potential energy landscape, which gradually changes its shape after every deformation step. 
Since we are looking for adjacent minima, line search algorithms, such as the steepest descent or conjugate gradient method, have proven effective in this regard. Because of its superior efficiency\reviewerone{ , in terms of number of iterations, which are costly in the described setup \cite{conjGradSteepDes},} we chose to use the conjugate gradient method for this task.

\subsection{The benchmark problem}
\label{sec:benchmark}
To demonstrate our procedure, we present the mechanical response of a binary Lennard-Jones glass \cite{Verlet1967, Tsamados2009LocalEM} to simple shear deformation. Glasses exhibit highly complex response behavior to mechanical deformation due to their inherently disordered structure, making response evaluation numerically challenging as it relies on fine local geometrical details in the atomic structure.

The Lennard Jones interatomic pair potential \cite{LJ_pot} is written as:
\begin{align}
    U(\bm{r}_{ij}) = 4\epsilon\left[\left(\frac{\sigma}{\bm{r}_{ij}}\right)^{12} - \left(\frac{\sigma}{\bm{r}_{ij}}\right)^{6}\right] \; ,
\end{align}
with $\sigma$ referring to the zero energy distances and $\epsilon$ quantifying the bond strength. The system consists of two different particle types, the small particle type (S) and the large particle type (L). The three zero-energy distance parameters are defined as:
\begin{align}
    \sigma_{SS} = 2\,\text{sin}\left(\frac{\pi}{10}\right) \; , \quad
    \sigma_{LL} = 2\,\text{sin}\left(\frac{\pi}{5}\right) \; , \quad
    \sigma_{SL} = 1 \; .
\end{align}
The three bond strength parameters are defined as:
\begin{align}
    \epsilon_{SS} = \frac{1}{2} \; , \quad
    \epsilon_{LL} = 1 \; , \quad
    \epsilon_{SL} = \frac{1}{2} \; .
\end{align}
We used Lennard-Jones units, and all output quantities depending on the potential are normalized with respect to a maximum value.

\begin{figure}
    \centering
    \begin{tikzpicture}
        \node at (0,0){\includegraphics[width=0.35\linewidth]{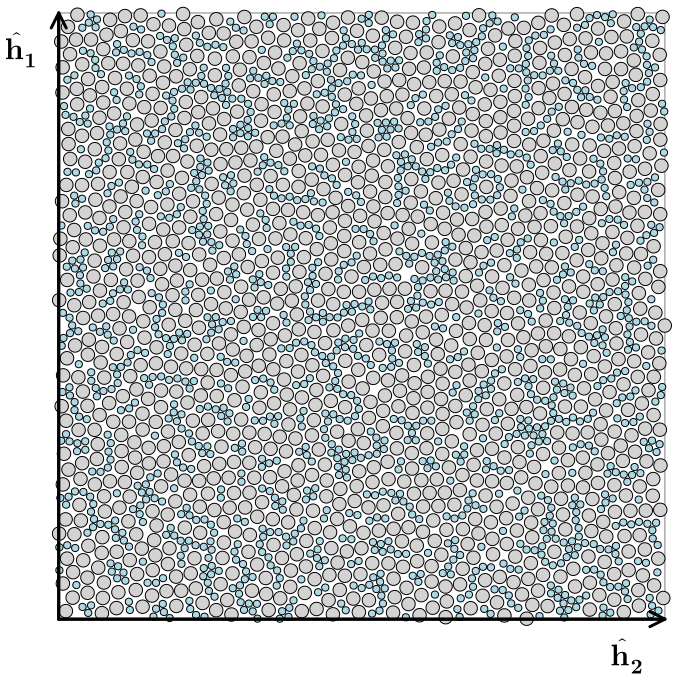}};
        \node at (7.5,0){\includegraphics[width=0.42\linewidth]{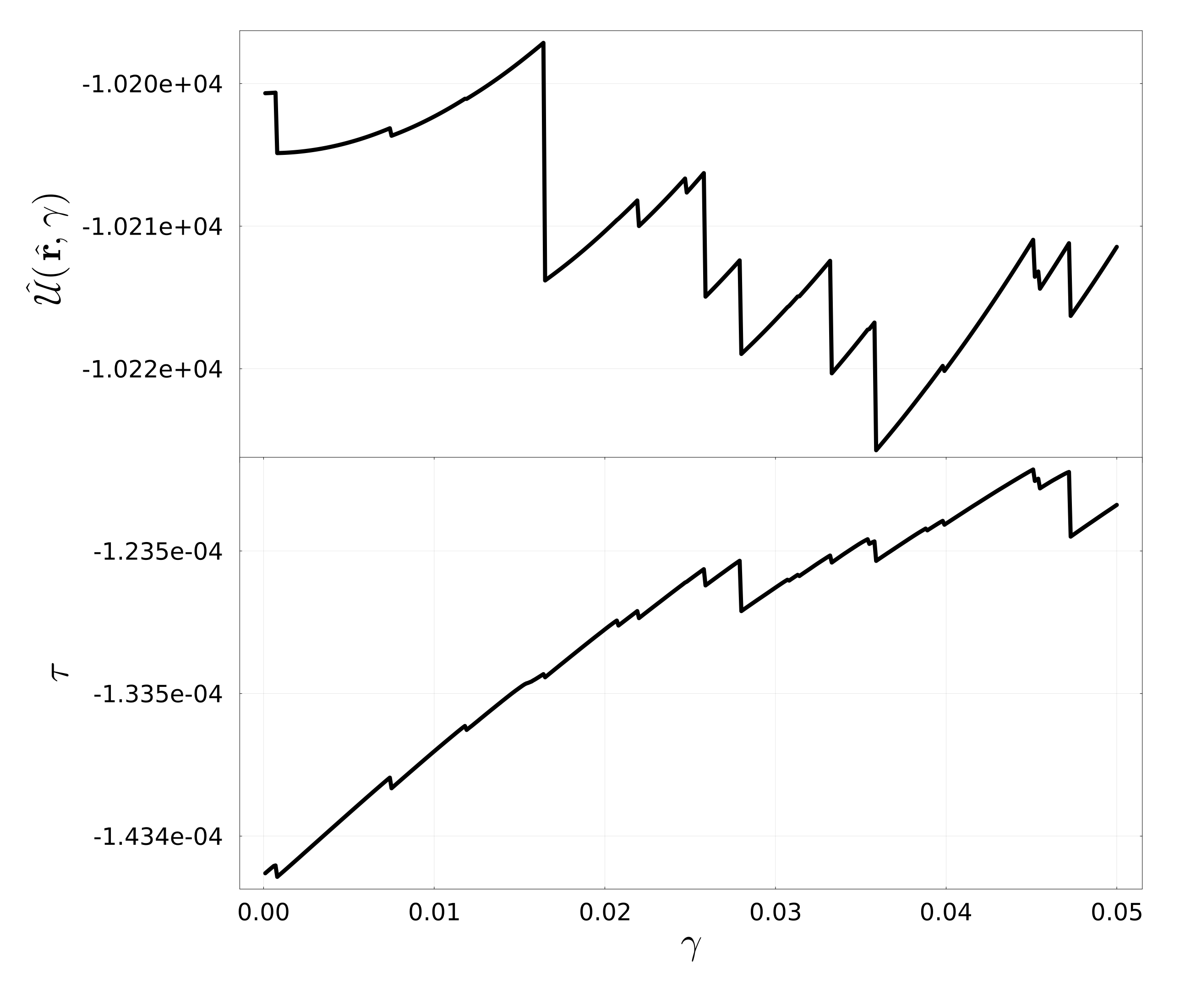}};
        \node at (0,3.25){\textbf{(a)}};
        \node at (7.5,3.25){\textbf{(b)}};
    \end{tikzpicture}
    \caption{The binary Lennard-Jones glass subjected to an athermal simple shear deformation protocol; (a) One out of $10^3$ material samples; (b) Potential energy $\hat{\mathcal{U}}(\hat{\bm{r}},\gamma)$, versus shear strain $\gamma$ and the corresponding stress-strain response $\tau:=\sigma_{12}$ versus shear strain $\gamma$.}
    \label{fig:benchmark_result}
\end{figure}

To rigorously test the new approach presented in this paper on a set of $10^3$ material samples, generated using the melting quenching technique \cite{frenkel2001molecularSim, Nos1986IsothermalisobaricCS}. This way, we first position $400$ atoms in a two-dimensional simulation cell. The length of the cell is chosen so that the material has an average density of one particle per \r A$^2$, leading to \reviewerone{square cells} with a side length of $20$ \r A in the initial configuration. The initial position is geometrically ordered, but the type of particle is random. This ensures physically possible states, prohibiting individual positions of the particles from being, by chance, too close. The ratio between \reviewerone{the number of} large and small particles is chosen as the golden mean, that is, $\frac{N_L}{N_S}=\frac{1+\sqrt{5}}{4}$ \cite{Jiang_2020}.
Melting quenching is performed within an NVT-ensemble \cite{frenkel2001molecularSim, Hoover1985, Nos1984AUF,Melchionna20021993}, where the number $N$ of particles, the volume $V$, and the temperature $T$ stay constant. This way, the $2N$-dimensional system is extended by an additional degree of freedom $\xi$, which allows for controlling the temperature of the system by manipulating the mean velocity of particles, acting as a thermostat to the ensemble \cite{frenkel2001molecularSim, allen2017, ditolla1993, Huenenberger2005}. The extended equations of motion were solved using the Velocity Verlet algorithm \cite{Verlet1967}.
The system was equilibrated for one million time integration steps above the melting temperature, using a step size of $0.005$ femtoseconds to ensure that no correlation between the initial and the equilibrated configuration was present. Starting from this equilibrated glass melt, we further equilibrated the system for $10^8$ steps, keeping the temperature constant and saving a total number of $10^3$ melted samples in equidistant time intervals of $10^6$ integration steps.
After that, every saved melted configuration was quenched for $10^6$ steps to $0.001$ Kelvin, allowing the system to approach a local minimum of the potential energy landscape. Finally, the conjugate gradient algorithm was performed to position the atomic structures precisely within the local basin of the potential energy landscape. This procedure results in $10^3$ uncorrelated, relaxed samples at zero temperature. One such sample is shown in Figure \ref{fig:benchmark_result}(a).
During the shearing process, meaningful information can be seen when the structure's potential energy and total stress are considered, as seen in Figure \ref{fig:schema_ex}.

Having generated the set of $10^3$ test samples, we subjected the configurations to \reviewerone{the AQS protocol using the simple shear}. The simple shear mapping during deformation increment $\Delta\gamma$ \reviewerone{for a single particle coordinates in two dimensions is defined as:}
\begin{align}\label{eq:disp_matrix}
    \bm{F}^{(\Delta \gamma)} = \left(
    \begin{matrix}
        1 & \Delta \gamma  \\
        0 & 1  
    \end{matrix} 
    \right) \in \mathbb{R}^{2 \times 2} \; .
\end{align}
Notably, the Bravais tensor is subjected to the simple shear mapping \reviewerone{$\bm{F}^{(\Delta \gamma)}$ as well}, while the particle configuration \reviewerone{vector $\bm{r}$ is} subjected to the mapping $\bm{F}^{(\Delta \gamma)}_N$  from \eqref{eq:Npower}. \reviewerone{The shear matrix in Equation \eqref{eq:disp_matrix} is just one example of a type of shear. The method is not limited to the simple shear and would also work for the pure shear (also known as pull-push shear), for which the shear matrix is defined as 
\begin{align}\label{eq:disp_matrix_pure_shear}
    \bm{F}^{(\Delta \gamma)} = \left(
    \begin{matrix}
        1 & \Delta \gamma_x  \\
        \Delta \gamma_y & 1  
    \end{matrix} 
    \right) \in \mathbb{R}^{2 \times 2} \; ,
\end{align}
where $\gamma_x$ and $\gamma_y$ are the pull and push parameters.
}

\section{\label{sec:SIP}Parallel stepping}
\subsection{Ansatz \label{sec:ansatz}}

By definition, every iteration during one deformation step \reviewerone{of the AQS procedure is formalized} by Equation \eqref{eq:AQS_iteration}, and depends on the previous \reviewerone{configuration $\bm{r}_{k-1}$}. As a consequence, before calculating the configuration $\bm{r}_k$, one needs to solve the \reviewerone{expensive} $ dN$-dimensional optimization problem $k-1$ times successively \reviewerone{before being able to calculate the configuration $\bm{r}_{k}$}. This can be extremely slow, as the deformation increments must be sufficiently small to 
\reviewerone{follow the minimum of the energy landscape accurately under deformation of the material.}
In the following, we present an approach, inspired by the \textit{Parareal} Algorithm \cite{LIONS2001661} for \reviewerone{ordinary} differential equations, offering a parallel computing scheme \reviewerone{for the AQS procedure.}

\reviewerone{\subsection{Level \rom{I} stepping \label{sec:lvl1stepping}}}
\reviewerone{Let $P\in\mathbb{N}_{>1}$ be equally capable threads (processors).}
The \reviewerone{proposed} procedure starts by performing a small \reviewerone{number $P$} of coarse steps, which we refer to as level \rom{I} stepping. \reviewerone{For this, we} introduce a coarse, \reviewerone{relatively large}, step size $\Delta \gamma^c=K\Delta\gamma$ for some $K\in\mathbb{N}_{>1}$, which, in turn, defines a new affine displacement matrix $\bm{F}^{(\Delta\gamma^c)}$, as shown in Equation \eqref{eq:disp_matrix} \reviewerone{for the case of simple shear}. 
\reviewerone{For a given configuration $\bm{r}_{\star}$ in its respective local energy minimum,} level \rom{I} stepping utilizes exclusively $P$ affine \reviewerone{displacements, which mathematically are given by a matrix vector multiplication using $F^{(i\Delta\gamma^c)}_N$ for $i=0,\dots,P-1$}.
\reviewerone{This induces a family of \reviewertwo{coarse} intermediate configurations as}
\begin{align} 
    \bm{r}_{iK}^c \coloneqq \bm{F}^{(i\Delta \gamma^c)}_N\bm{r}_{\star}\qquad \forall i=0,\ldots,P-1.
\end{align}
\reviewerone{Note that no minimization is performed for any member of the family. 
Further, for $i=0$, we observe that $\bm{r}_{\star}=\bm{r}_{0}^c$, and for $i > 0$ the configurations of this family are not located} in \reviewerone{their respective local} minima of the potential energy landscape. 
Notably, this sequence of affine mappings is computationally insignificant since these mappings are realized by \reviewerone{simple} matrix-vector multiplications. 
This level \rom{I} procedure provides a set of \reviewerone{intermediate initial configurations $\bm{r}_{iK}^c$,} each of which serves as a starting \reviewerone{configuration} for further processing \reviewerone{on each of the individual threads}.

\reviewerone{\subsection{Level \rom{II} stepping \label{sec:lvl2stepping}}}
\reviewerone{Every thread during level \rom{II} stepping first minimizes (in parallel) the intermediate initial configurations $\bm{r}_{iK}^c$ from level \rom{I} to descent to its corresponding minimum $\mathcal{F}(\bm{r}_{iK}^c)$.
Then, starting from $\mathcal{F}(\bm{r}_{iK}^c)$, each thread performs athermal quasistatic (AQS) deformation steps individually, proposing a sequence of equilibrium candidates. 
Mathematically, this consists of applying the operator $\mathcal{F}\circ F^{(\Delta\gamma)}$ $K$-times  to $\mathcal{F}(\bm{r}_{iK}^c)$ for all $i=0,\ldots,P-1$ in parallel on the corresponding threads.
This results in the family of equilibrium candidates corresponding to the shear value $k\Delta\gamma$ relative to the shear value of $\bm{r}_{\star}$
\begin{align}\label{eq:cond_notation}
    \bm{r}_{k}|\bm{r}_{iK}^c \coloneqq \left(\left(\mathcal{F}\circ \bm{F}^{(\Delta \gamma)}_N\right)^{k-iK} \circ \mathcal{F}\right)(\bm{r}_{iK}^c) = \left( \left(\mathcal{F}\circ \bm{F}^{(\Delta \gamma)}_N\right)^{k-iK} \circ \mathcal{F} \circ \bm{F}^{(i\Delta\gamma^c)}_N\right)(\bm{r}_{\star}),
\end{align}
for all $k=iK,\ldots, iK+K$ and $i=0,\dots,P-1$. As one can see from the indices of Equation \eqref{eq:cond_notation} there can be multiple candidates for the same shear value from different threads. 
To be precise, any index $\hat{k}\in K\mathbb N$ induces a configuration for a shear value $\hat{k}\Delta \gamma$ consisting of two different candidates. Namely, the \textbf{outgoing} configuration $\bm{r}_{\hat{k}}|\bm{r}^c_{\hat{k}}$ and \textbf{the incoming} configuration $\bm{r}_{\hat{k}}|\bm{r}^c_{\hat{k}-K}$. 
\reviewerone{In other words}, starting at a particular level \rom{I} initial configuration $\bm{r}_{iK}^c$ and performing $K$ level \rom{II} fine AQS steps to obtain the incoming configuration $\bm{r}_{(i+1)K} | \bm{r}^c_{iK}$ corresponds to the same shear as the subsequent level \rom{I} coarse configuration $\bm{r}_{(i+1)K}^c$ that yields, after minimization, the outgoing configuration $\bm{r}_{(i+1)K} | \bm{r}^c_{(i+1)K}=\mathcal{F}\left(\bm{r}_{(i+1)K}^c\right)$.
}
Those configurations may not be equal in the sense of being in the same local basin of attraction due to the non-convex nature of the potential energy landscape. 

\begin{figure}[htbp]
    \centering
     \includegraphics[width=1.0\linewidth]{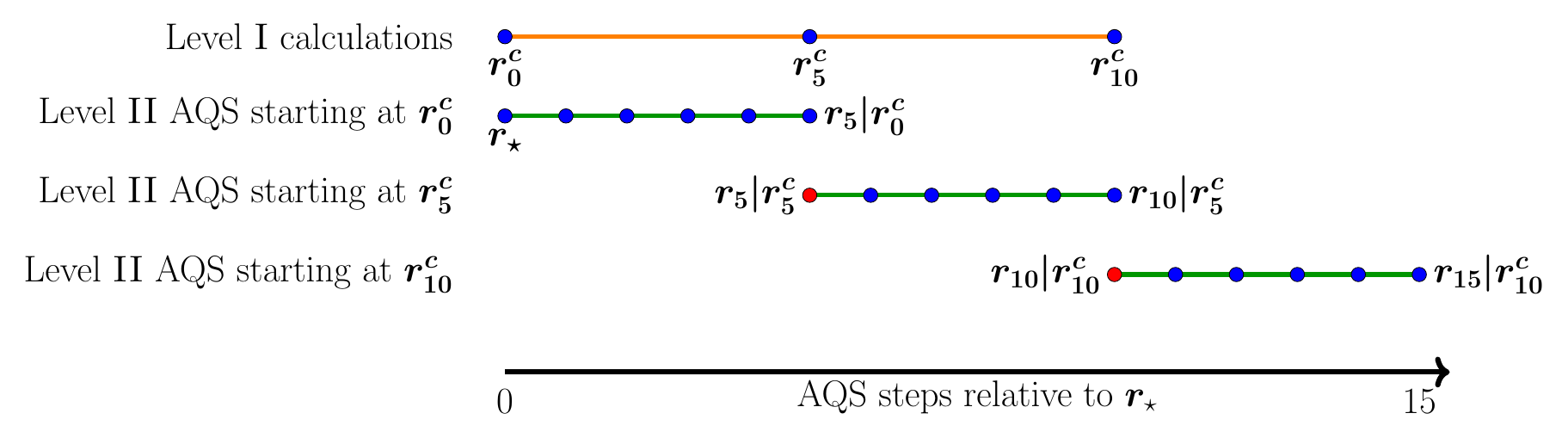}
    \caption{\reviewerone{Scheme of parallel computing approach. The first row is calculated sequentially, while the other three rows are done in parallel. After the AQS procedure on every thread the results are checked for equality on relevant indices.}}
    \label{fig:aqs_sip}
\end{figure}

\reviewerone{
\subsection{Equality checking \label{sec:equality}}}

In this section, we precisely define the concept of equality within our framework and establish what constitutes an accepted---or equivalently, correct---solution in the simulation. Two configurations $\bm{r}_a, \bm{r}_b\in \conv(\bm{H}^{(\gamma)})^N$ are considered equal if and only if they reside within the same basin of attraction defined in \eqref{eq:basin_of_attr}:
\begin{align}
    \label{eq:local_min}
    \bm{r}_a \stackrel{\sim}{=} \bm{r}_b \qquad \Leftrightarrow \qquad \bm{r}_a,\bm{r}_b \; \in\;
    \Omega_{\bm{r}_{\infty},\gamma}\;.
\end{align}

\reviewerone{Then, a configuration $\bm{r}_{\square}$ is considered correct if \textbf{at least one} of the following two conditions is fulfilled:
\begin{enumerate}
    \item \label{enum:correct_config} It originates from a correct configuration $\bm{r}_{\circ}$, meaning an arbitrary positive number $m > 0$ of fine steps are performed in between, i.e.  $\bm{r}_{\square}=(\mathcal{F}\circ \bm{F}^{(\Delta \gamma)}_N)^m \;(\bm{r}_{\circ})$.
    \item \label{enum:equality} There exists a correct configuration $\bm{r}_{\circ}$ corresponding to the same shear value with 
    \[\bm{r}_{\square} \quad \stackrel{\sim}{=} \quad \bm{r}_{\circ}.\] 
\end{enumerate}
The latter condition \ref{enum:equality} is particularly important if 
the correct configuration is an incoming configuration e.g. $\bm{r}_{\circ}\;\stackrel{\sim}{=}\;\bm{r}_{iK}|\bm{r}_{(i-1)K}^c$ and the configuration considered is an outgoing configuration $\bm{r}_{\square}\;\stackrel{\sim}{=}\;\bm{r}_{iK}|\bm{r}_{iK}^c$. Then, if $$\bm{r}_{iK}|\bm{r}_{iK}^c \quad\stackrel{\sim}{=} \quad\bm{r}_{iK}|\bm{r}_{(i-1)K}^c$$ holds, all subsequent configurations calculated via the AQS procedure starting at the outgoing configuration $\bm{r}_{iK}|\bm{r}_{iK}^c$ can be considered correct using condition \ref{enum:correct_config} for some $i=1,\dots,P-1$. Then, we can accept all configurations up to $\bm{r}_{(i+1)K}|\bm{r}_{iK}^c$ as correct.} In the following, we introduce the implementation of the parallel stepping scheme, taking into account the correction of possible miss-predictions during level \rom{II} stepping \reviewerone{based on incorrect outgoing configurations.}

\reviewerone{\subsection{Parallel calculation scheme \label{sec:calculation_scheme}}}
\reviewerone{The proposed parallel procedure operates through a predictor-corrector iterative framework designed to accelerate the AQS procedure. The core algorithmic loop can be summarized by the following principal steps and its pseudo code is provided in Algorithm \ref{alg:parallel}:

\begin{enumerate}
    \item \textbf{Initialization:} The procedure begins by initializing the reference state as the initial configuration, setting $\bm{r}_{\star} \gets \bm{r}_0$. We consider the initial configuration to be correct.
    
    \item \textbf{Two-Level Evolution:} During each main iteration (inside the While loop in Algorithm \ref{alg:parallel}) the scheme executes evaluations on two levels. First, level \rom{I}  generates a set of $P$ intermediate initial configurations $\bm{r}_{iK}^c$ indexed by $i$. Level \rom{II} then performs a fine simulation on these intermediate initial configurations $\bm{r}_{iK}^c$ to compute the subsequent results in parallel.
    
    \item \textbf{Equality check and update:} A sequential verification step is performed to compare the results of adjacent index intervals $\hat{k}\in K\mathbb{N}$. The relevant configurations are the respective incoming and outgoing configurations for each $i=1,\dots,P-1$. If the correctness condition \ref{enum:equality}, formally $\mathbf{r}_{iK} | \mathbf{r}_{(i-1)K}^c \; \stackrel{\sim}{=}\; \mathbf{r}_{iK} | \mathbf{r}_{iK}^c$, holds across all threads, the reference state is updated to the final computed result $\mathbf{r}_{\star} \gets \mathbf{r}_{P\cdot K}$. Conversely, if a miss-prediction is detected, the algorithm identifies the highest index $q$ for which we can consider the corresponding configuration $\bm{r}_q$ as correct, effectively rejecting all subsequent steps calculated. The reference state is then updated $\mathbf{r}_{\star} \gets \mathbf{r}_q$.
    \item \textbf{Iteration:} 
    The two-level evolution cycle (steps 2 and 3) is repeated until the total target number of steps $N_{steps}$ is achieved.
\end{enumerate}
}

Let's take a look at Figure \ref{fig:aqs_sip}, where we schematically perform a $15$-step AQS procedure applying parallel stepping. Using three threads for $15$ steps in total, one defines three level \rom{I} \reviewerone{intermediate} starting configurations \reviewerone{$\left\{\bm{r}_{0}^c,\bm{r}_5^c,\bm{r}_{10}^c\right\}$ by multiplying $F_N^{(i\Delta\gamma^c)}$ onto $\bm{r}_{\star}$ for $i=0,1,2$}.
This way, one initiates three individual level \rom{II} AQS stepping schemes on three threads, simultaneously starting from \reviewerone{$\bm{r}_{0}^c$, $\bm{r}_5^c$}, and $\bm{r}_{10}^c$ respectively. After this, the first set of solution candidates for the AQS response trajectory is set up to step $15$. We now \reviewerone{recognize} incoming configurations as \reviewerone{$\bm{r}_5|\bm{r}_{0}^c$} and $\bm{r}_{10}|\bm{r}^c_{5}$ and outgoing configurations as $\bm{r}_5|\bm{r}_5^c$ and $\bm{r}_{10}|\bm{r}_{10}^c$ for the indices $5$ and $10$, respectively. Notably $\bm{r}_5|\bm{r}_5^c$ and $\bm{r}_{10}|\bm{r}_{10}^c$ are the results of the minimization operation $\mathcal{F}$ on $\bm{r}_5^c$ and on $\bm{r}_{10}^c$, respectively. In the following, we check the validity of the solution candidates and \reviewerone{reject} them if necessary.
This means that, if $\bm{r}_5$ and $\bm{r}_5|\bm{r}_5^c$ are equal, then the solution $\bm{r}_{10}|\bm{r}_5^c$ is also correct, however, if $\bm{r}_5$ and $\bm{r}_5|\bm{r}_5^c$ are not equal we must initiate the procedure again \reviewerone{by setting $\bm{r}_{\star}\leftarrow \bm{r}_5$}, while rejecting all \reviewerone{solution candidates based on intermediate configurations $\bm{r}_5^c$ and $\bm{r}_{10}^c$}. \reviewerone{In  reference to Algorithm \ref{alg:parallel} this is done by executing lines 5 to 16.}

\begin{figure}[htbp]
    \centering
    \includegraphics[width=1.0\linewidth]{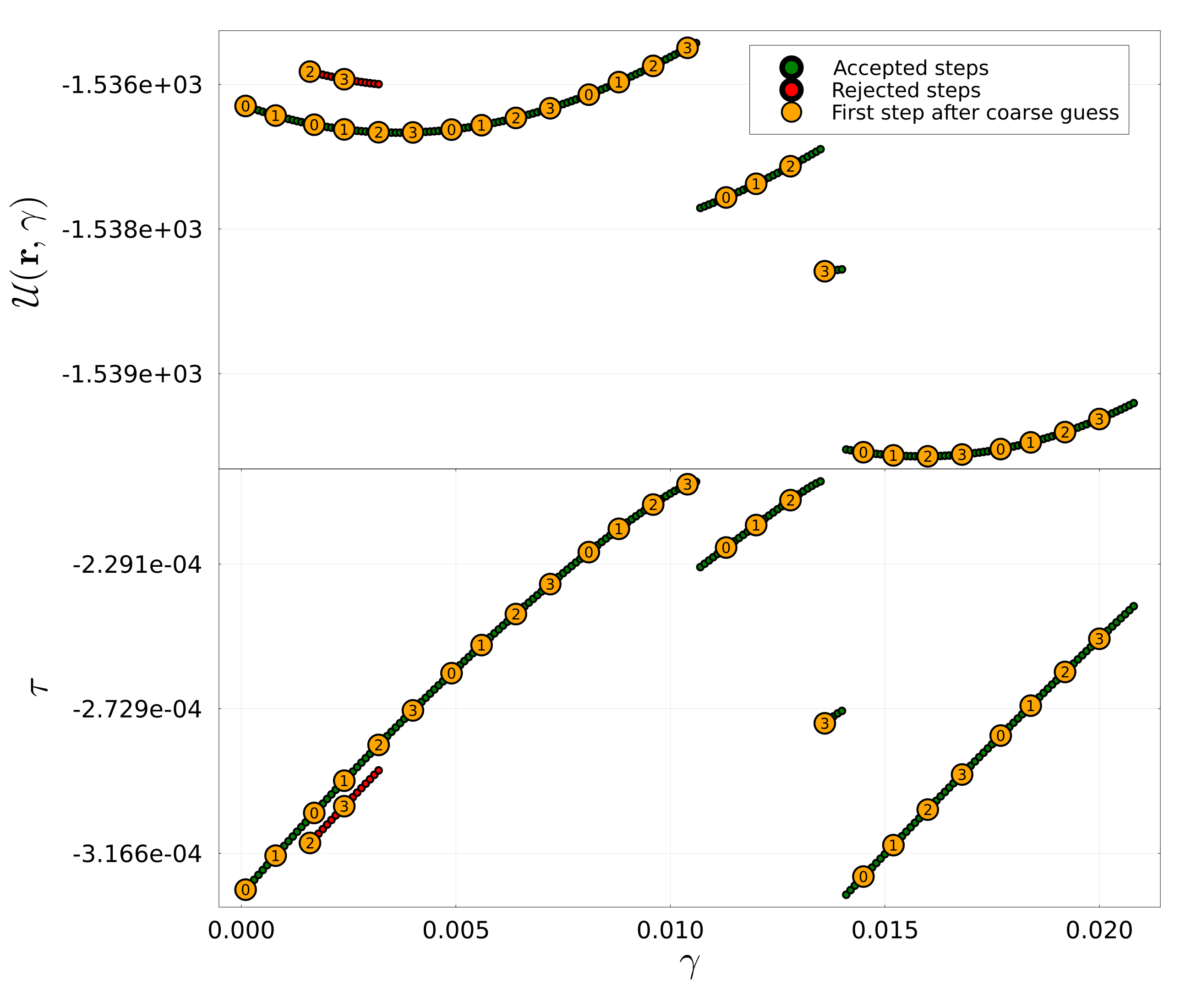}
    \caption{Example AQS simulation in parallel scheme for $P=4$ and a total of $200$ steps with $\Delta\gamma=\frac{1}{200}$. On each thread, $K=8$ fine steps were performed per iteration. The orange dots represent the node points, each is marked with the thread index. If the following \textit{outgoing configuration} is equal, the steps are accepted and the points are colored in green. If a configuration is not accepted at the node, the configurations that follow are colored red.}
    \label{fig:schema_ex}
\end{figure}

In Figure \ref{fig:schema_ex}, we \reviewerone{ show the proposed method on a real example where} one can see that the method can work in parallel, successfully reducing the total computational time.

\reviewerone{
\begin{algorithm}
\caption{Parallel procedure}
\label{alg:parallel}
\begin{algorithmic}[1]
\Procedure{aqs\_parallel}{$\bm{r}_0,\; \hat{\bm{H}},\; F^{(\cdot)}_N ,\; \Delta \gamma,\; \Delta\gamma^c,\; N_{steps},\; K,\; P$}
    \State $N_{current}\gets 0$
    \State $\bm{r}_{\star}\gets \bm{r}_0$
    \While{$N_{current}<N_{steps}$}
        \State  Level \rom{I} steps: $\{ \bm{r}^c_{iK} \;\gets \; \;\bm{F}_N^{(i \,\Delta\gamma^c)} \; \left(\bm{r}_{\star}\right)\; |\; i=0,\dots,P-1 \; \}$ \Comment{Intermediate initial configurations}
        \For{$i = 0$ \textbf{to} $P-1$} \Comment{Parallel procedure}
            \State Level \rom{II} steps: $\bm{r}_{(i+1)K}\,|\,\bm{r}_{iK}^c \; \gets\; \left(\mathcal{F}\circ\bm{F}^{(\Delta \gamma)}_N\right)^K \circ\mathcal{F}\circ(\bm{r}_{iK}^c)$
        \EndFor
        \For{$i = 1$ \textbf{to} $P-1$} \Comment{Sequential equality check}
            \If{$\bm{r}_{iK}\,|\,\bm{r}_{(i-1)K}^c \stackrel{\sim}{=} \bm{r}_{iK}\,|\,\bm{r}_{iK}^c$}
                \State $q \gets (i+1)K$
                \State continue
            \Else
                \State $q \gets iK$
                \State break \textbf{for}
            \EndIf
        \EndFor
        \State $\bm{r}_{\star} \gets \bm{r}_q$
        \State $N_{current} \gets N_{current} + q$
    \EndWhile
\EndProcedure
\end{algorithmic}
\end{algorithm}
}

\subsection{Response trajectories in the potential energy landscape}

\begin{figure}
    \centering
    \includegraphics{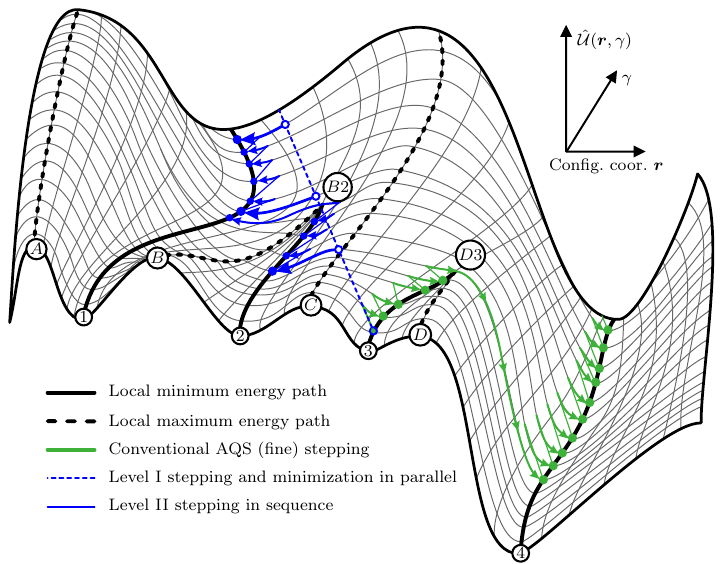}
    \caption{Evolution of a one-dimensional potential energy landscape with increasing shear deformation.}
    \label{fig:pel_sketch}
\end{figure}

In the following, we will break down \reviewerone{the AQS procedure and the parallel method} with the help of the visualization of the evolution of a one-dimensional potential energy landscape varying over the shear strain $\gamma$, shown in Figure \ref{fig:pel_sketch}. In this figure, the evolution of local minima is presented by black \reviewertwo{dashed} lines, while the evolution of local maxima is presented by \reviewertwo{solid} black lines. This illustrative example reveals four maxima \reviewerone{paths}, referred to as $A$, $B$, $C$, and $D$, and four minima \reviewerone{paths}, referred to as $1$, $2$, $3$, and $4$.
Our solution trajectory candidates lie on local minimum energy paths. Minimum and maximum energy paths may merge into saddle points when increasing the shear strain, where the configuration jumps from one to an adjacent equilibrium state. In Figure \ref{fig:pel_sketch}, we see one saddle node when maximum $B$ and minimum $2$ merge and another saddle node when maximum $D$ and minimum $3$ merge. With increasing strain, the curvature gradually decreases, which can be quantified by the lowest eigenvalue of the Hessian of the potential energy landscape, until the minimum merges into the saddle node, the \reviewerone{equilibrium} becomes unstable. \reviewerone{The lowest eigenvalue also corresponds to the smallest eigenvalue as the Hessian of this system is symmetric and semi-positive definite.} During such an event, the potential energy drops, leading to a sudden finite change in the configurational coordinate $\bm{r}$. In the configuration space, such events are localized sudden atomic-scale rearrangements while the surrounding material responds elastically, often also referred to as shear transformations \cite{bamer2023}.
Let us assume now that a minimum of $3$ is the correct value at zero shear strain and \reviewerone{let us }proceed with the exploration from there. A conventional step-by-step AQS protocol is presented in green in Figure \ref{fig:pel_sketch}. Proceeding in finite strain increments, one macroscopically deforms the system and pushes the configuration out of minimum $3$. Every minimization step pushes the system back into the minimum configuration while the shear strain is \reviewerone{constant}. The system is further driven to the saddle node $D3$ and drops down into the adjacent minimum $4$, where the system is further driven by external shear.\\
The coarse level \rom{I} stepping is represented in blue in Figure \ref{fig:pel_sketch}. Here, we show a scenario in which the parallel stepping scheme \reviewerone{leads} to an incorrect response trajectory and will be rejected. One starts at the same point and pushes the system into the same affine direction as the conventional stepping scheme. We remind that level \rom{I} stepping is a sequence of exclusively affine mappings, which means we receive the intermediate deformation states indicated by the white markers on an extension of the same line. We see that, performing a larger stepping length, the configuration is pushed across the maximum energy path $C$ already after the first affine mapping, and it remains on the other side of the local maximum energy path $C$. However, one does not know that as long as one does not evaluate the corresponding inherent states by performing minimization of the potential energy landscape.\\ As more elaborately discussed in Sections \ref{sec:ansatz}, \ref{sec:lvl2stepping} and \ref{sec:equality}, we perform now level \rom{II} stepping, which consists of initiating multiple AQS sequences of fine stepping from the configurations indicated by the blue markers in Figure \ref{fig:pel_sketch}. Notably, one starting point is located on one side of the local maximum energy path $C$ while the other three starting points are located on the other side of $C$. However, only after checking different states at the same macroscopic strain for equality according to Section \ref{sec:equality}, we can reject the level \rom{I} proposals. Considering the special case in Figure \ref{fig:pel_sketch}, the incoming configuration is indicated in magenta in the local minimum energy path $4$ while the corresponding outgoing configuration is indicated in magenta in the local minimum energy path $2$. Consequently, all strain states after the magenta configurations are not considered further, and new level \rom{I} proposals are set from there.
We note that the solution trajectory of stepping is numerically equivalent to the one obtained by performing conventional AQS deformation stepping. However, what we generally observe when performing \reviewerone{the} parallel stepping \reviewerone{method} is the possibility of getting trapped in alternative solution paths, eventually leading to rejections and recalculations. This occurs when the level one stepping crosses local maxima and explores new basins in the potential energy landscape. Since the problem is highly nonlinear in its nature, proceeding along an alternative minimum energy path without rejection could lead to an entirely different macroscopic failure picture in the material.

Athermal stepping is purely deterministic \reviewerone{in the sense of adiabatic continuation}. Thus, repeating one simulation will result in exactly the same outcome. However, this is not the case when including temperature \reviewerone{as one would do in a more classical molecular dynamics framework}. Thus, from a molecular dynamics perspective, \reviewerone{overcoming an energy barrier,} such as path $C$ in Figure \ref{fig:pel_sketch}, \reviewerone{to get to another basin of attraction,} is a scenario that might occur during thermal deformation since thermal vibration may result in the system falling into the minimum energy path $2$, \reviewerone{which would not be the case in a purely athermal setting}. Clearly, the probability of this happening depends largely on the activation energy necessary to reach the maximum energy \reviewerone{configuration} in $C$ and the height of the opposite maximum energy \reviewerone{configuration} $D$. Therefore, although proceeding along path $1$ in Figure \ref{fig:pel_sketch} is strictly speaking incorrect from a mathematical perspective, it may constitute a physically meaningful response trajectory and might be valuable for investigations at finite temperature. \reviewerone{This is because the configuration is still in the respective local energy minimum but there is no adiabatic continuation when considering the starting configuration was at $3$ and not $1$ in Figure \ref{fig:pel_sketch}.}

\begin{figure}[htbp]
    \centering
    \includegraphics[width=1.0\linewidth]{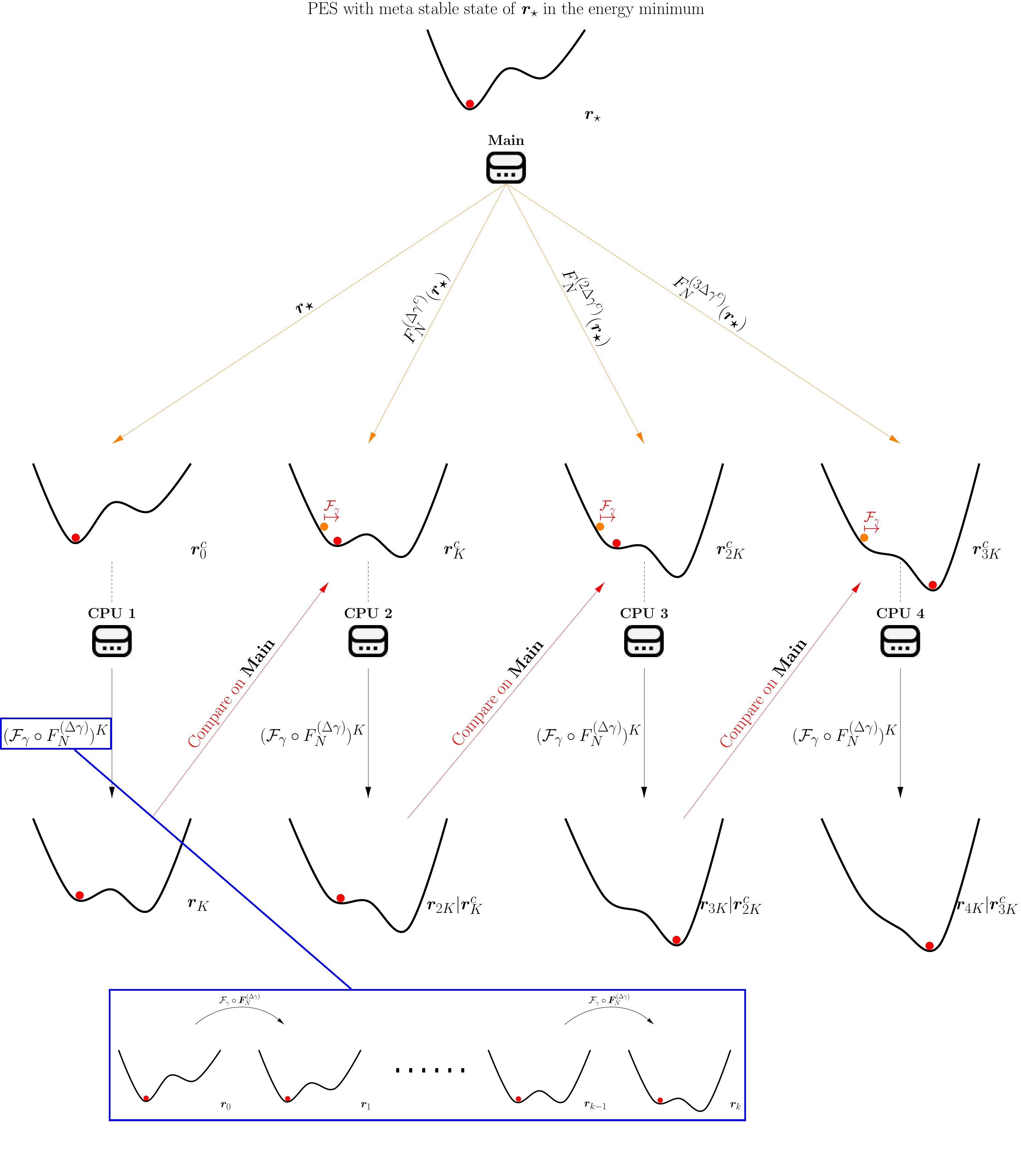}
    \caption{\reviewerone{Scheme of parallel computing approach. The intermediate steps are distributed to the different threads on which the fine steps are calculated. On every thread a $K$-step AQS procedure is simulated as illustrated with Figure \ref{fig:aqs_protocol}(d) in the blue box.}}
    \label{fig:aqs_sip2}
\end{figure}

\reviewerone{
\subsection{Finite termination and speed-up bounds}
In this section we discuss what the best and worst case scenarios in terms of speed-up are and what guarantees one can expect. We consider the standard sequential AQS procedure as our baseline. Since we know the initialization step $\bm{r}_{\star}$ to be correct in the sense of \ref{enum:equality}, the first thread in the sequence $i=0$ will always compute the exact fine sequential response for its assigned interval of $K$ steps. Depending on the topography of the potential energy landscape and the accuracy after the level \rom{I}  stepping, the algorithm's performance is strictly bounded by the following best-case and worst-case scenarios.

\subsubsection*{Best-Case Scenario: Ideal Convergence and Maximum Speed-Up}
\begin{itemize}
    \item \textbf{Mechanism:} The optimal scenario occurs when the computationally insignificant sequence of affine mappings during level \rom{I} stepping, defined by $\bm{r}_{iK}^c := \bm{F}_N^{(i\Delta\gamma^c)} \bm{r}_{\star}$, perfectly predicts the correct local basin of attraction for every intermediate initial configuration for level \rom{II} on every thread.
    \item \textbf{Speed-Up Bound:} Because the correctness condition is fulfilled across all threads, the reference state is immediately updated to the final computed result of the iteration. The algorithm successfully accepts $P \cdot K$ fine steps simultaneously. Consequently, the theoretical speed-up approaches the total number of equally capable threads, $P$, bounded only by the negligible overhead of the level \rom{I} matrix-vector multiplications.
\end{itemize}

\subsubsection*{Worst-Case Scenario: Sequential Fallback and Zero Speed-Up}
\begin{itemize}
    \item \textbf{Mechanism:} 
    The worst-case limit manifests when at each parallel iteration (lines 5-19 Algorithm \ref{alg:parallel}), the second thread $i=1$ of the parallel level \rom{I} provides $\bm{r}_{K}^c$ which resides in the wrong basin of attraction and, subsequently, the equality check $\bm{r}_{K}|\bm{r}_{K}^c \stackrel{\sim}{=} \bm{r}_{K}|\bm{r}_{0}^c$ fails. 
    \item \textbf{Finite Termination:} Even under continuous prediction failure, the algorithm is strictly guaranteed to advance. The thread handling $i=0$ initiates its level \rom{II} AQS stepping from the correct, verified reference state $\bm{r}_{\star}$. After performing $K$ fine steps, the reference state is updated ($\bm{r}_{\star} \gets \bm{r}_q = \bm{r}_K$) to this newly verified configuration, locking in $K$ correct steps before the next parallel iteration begins and the method advances by one interval of $K$ fine steps.
    \item \textbf{Speed-Up Bound:} By advancing at least $K$ steps per iteration, the algorithm deterministically reaches the total target number of steps, $N_{steps}$, in a finite bound of exactly $N_{steps}/K$ parallel iterations. In this worst-case scenario, the computational work of threads $i=1, \dots, P-1$ is entirely discarded, resulting in a speed-up of $1$ (or slightly less, due to the overhead of rejected parallel evaluations), effectively reducing the algorithm to the conventional step-by-step AQS protocol.
\end{itemize}
}

\reviewerone{\section{Numerical results}}
\label{sec:results}
\reviewerone{We have performed the AQS procedure on $1000$ different samples}. All $1000$ \reviewerone{sample initial configurations}\reviewertwo{, which have a pseudo-randomized structure and use the same setup as the benchmark problem in Section \ref{sec:benchmark},} have been deformed using the standard AQS method and the \reviewerone{\NewAnswers{newly proposed parallel method}}. For all samples, the same computer architecture and \reviewerone{\NewAnswers{shear step size}} $\Delta\gamma=\frac{1}{200}$ with $N_{steps}=200$ was used, while changing the number of threads $P$ and \reviewerone{\NewAnswers{the parameter}} $K$. 
Notably, all \reviewerone{final configurations} were equal in the sense defined in Section \ref{sec:equality}.
\\

\begin{figure}[htbp]
    \centering
    
    \begin{subfigure}[b]{0.49\textwidth}
        \centering
        \includegraphics[width=\textwidth]{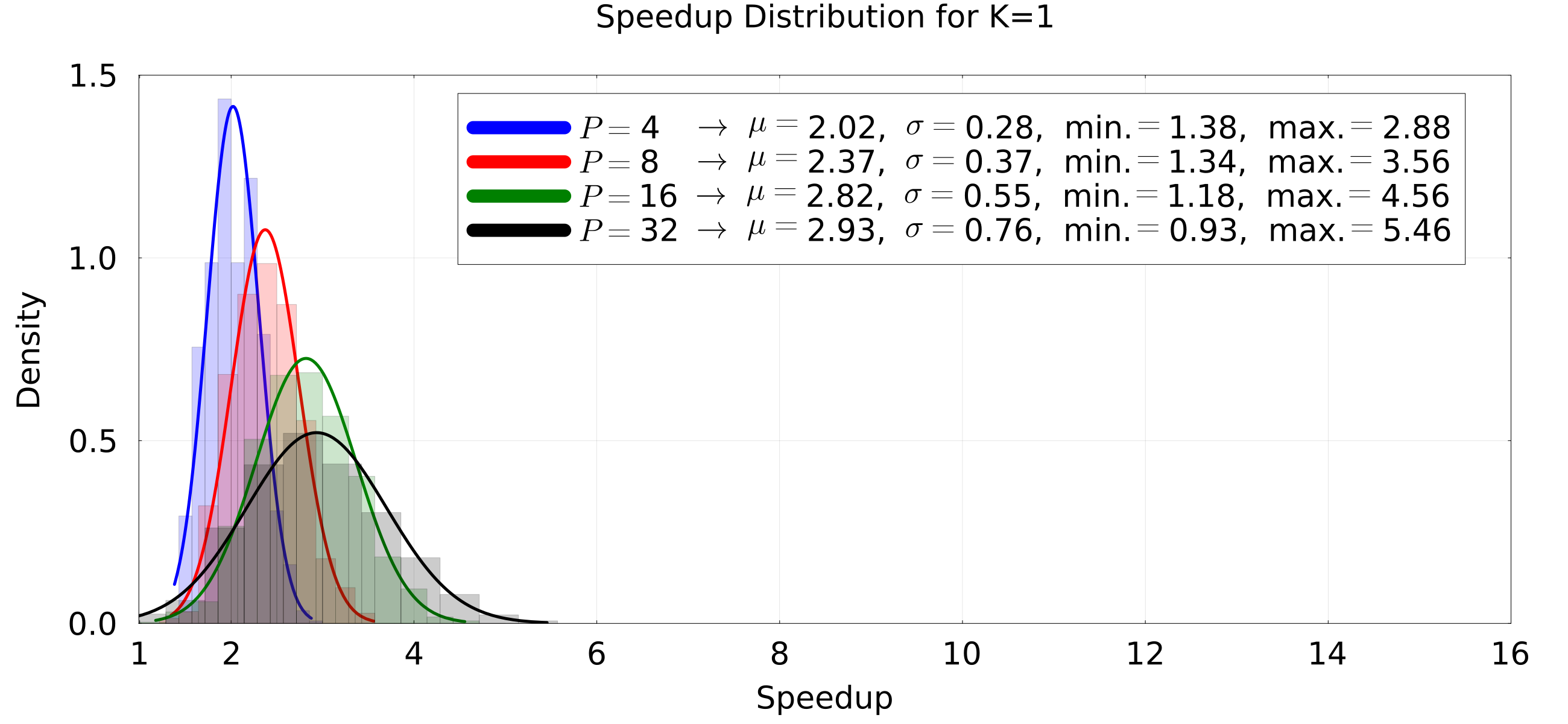}
        \label{fig:sub_a}
    \end{subfigure}
    \hfill 
    \begin{subfigure}[b]{0.49\textwidth}
        \centering
        \includegraphics[width=\textwidth]{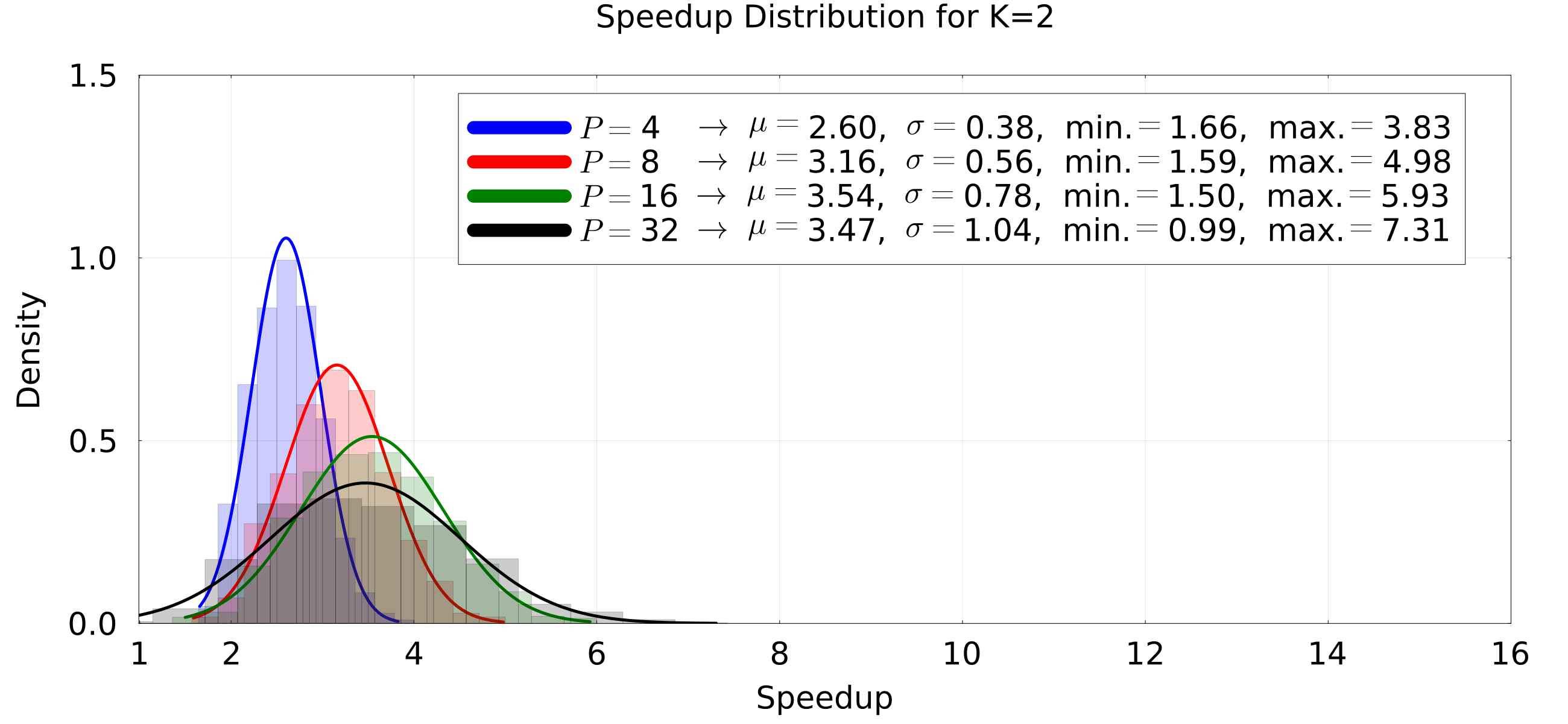}
        \label{fig:sub_b}
    \end{subfigure}
    
    \vspace{1em}
        \begin{subfigure}[b]{0.49\textwidth}
        \centering
        \includegraphics[width=\textwidth]{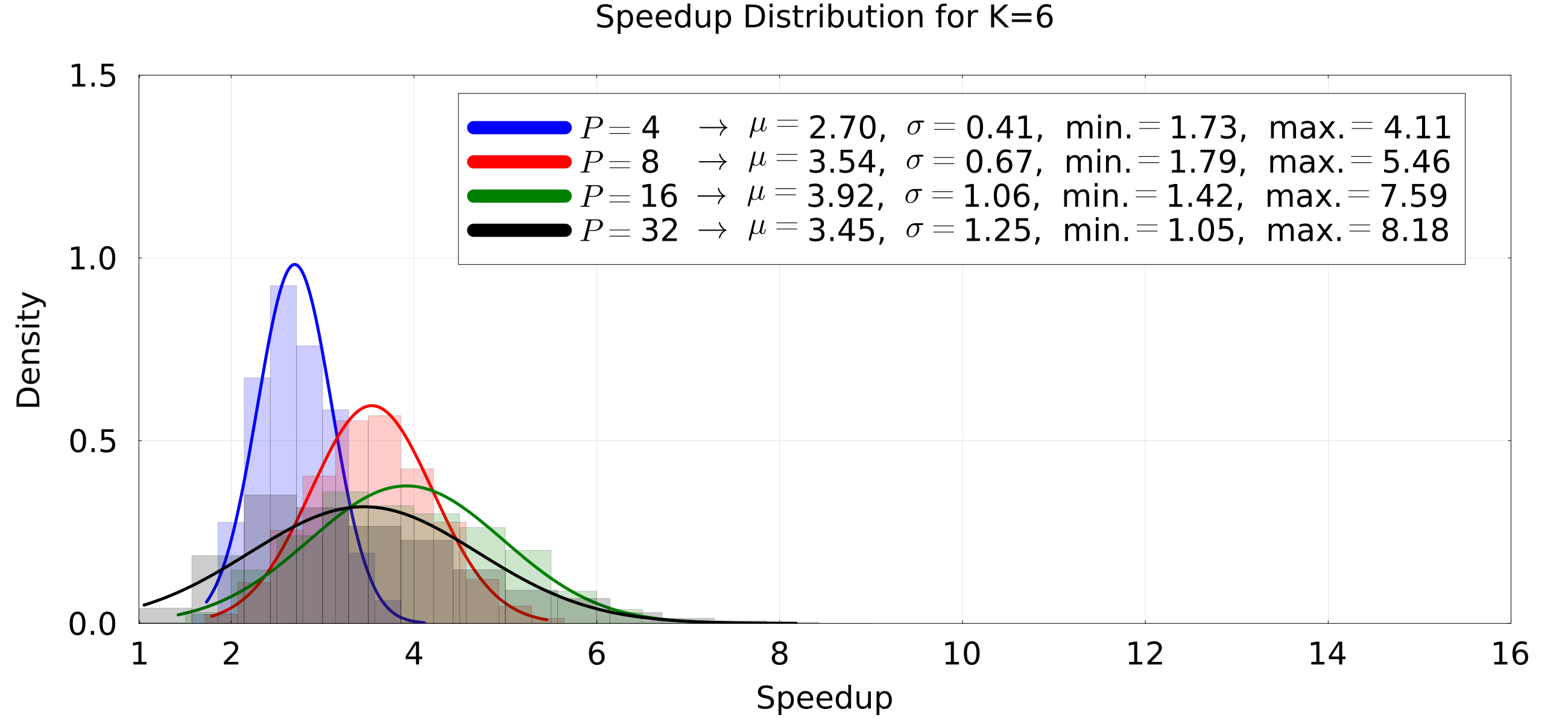}
        \label{fig:sub_c}
    \end{subfigure}
    \hfill 
    \begin{subfigure}[b]{0.49\textwidth}
        \centering
        \includegraphics[width=\textwidth]{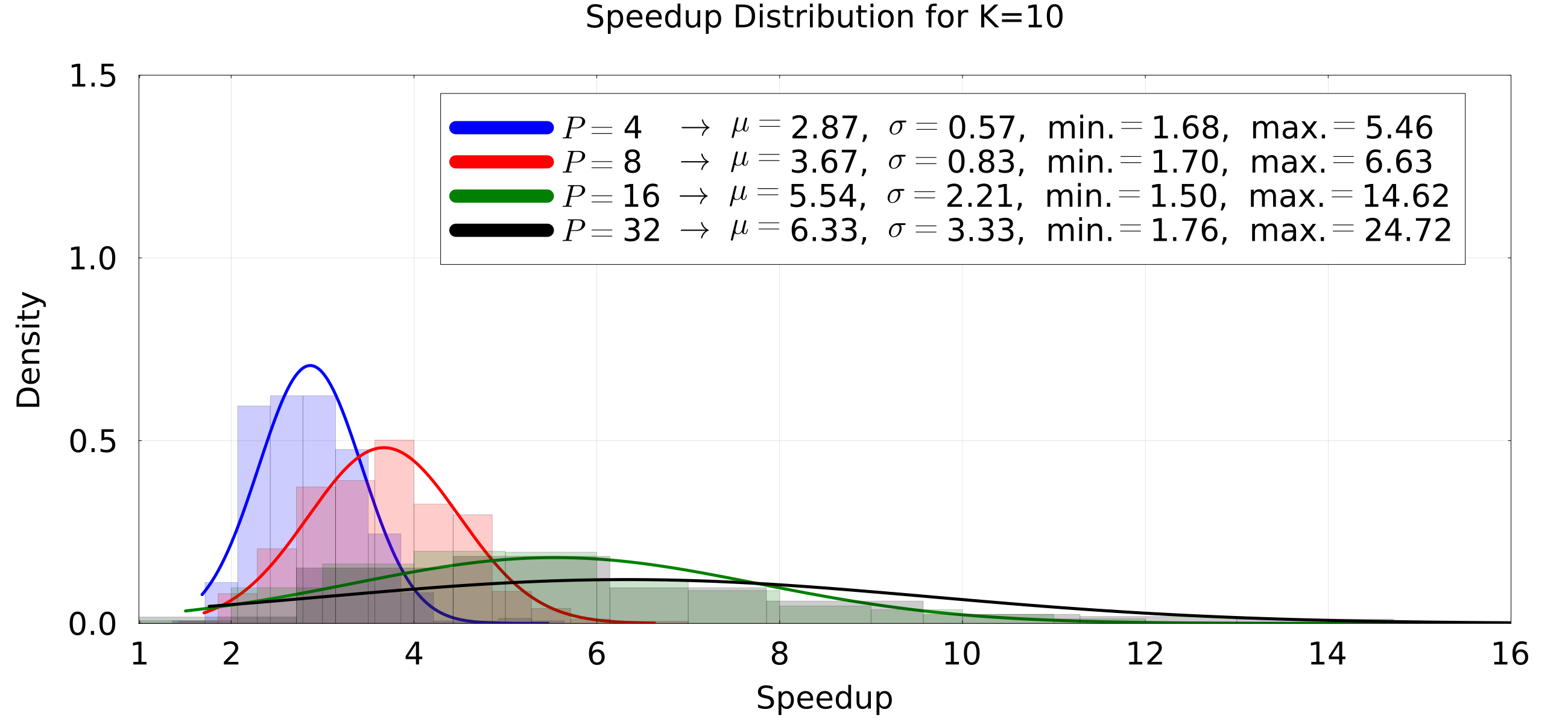}
        \label{fig:sub_d}
    \end{subfigure}
    
    \vspace{1em}
    
    \begin{subfigure}[b]{0.49\textwidth}
        \centering
        \includegraphics[width=\textwidth]{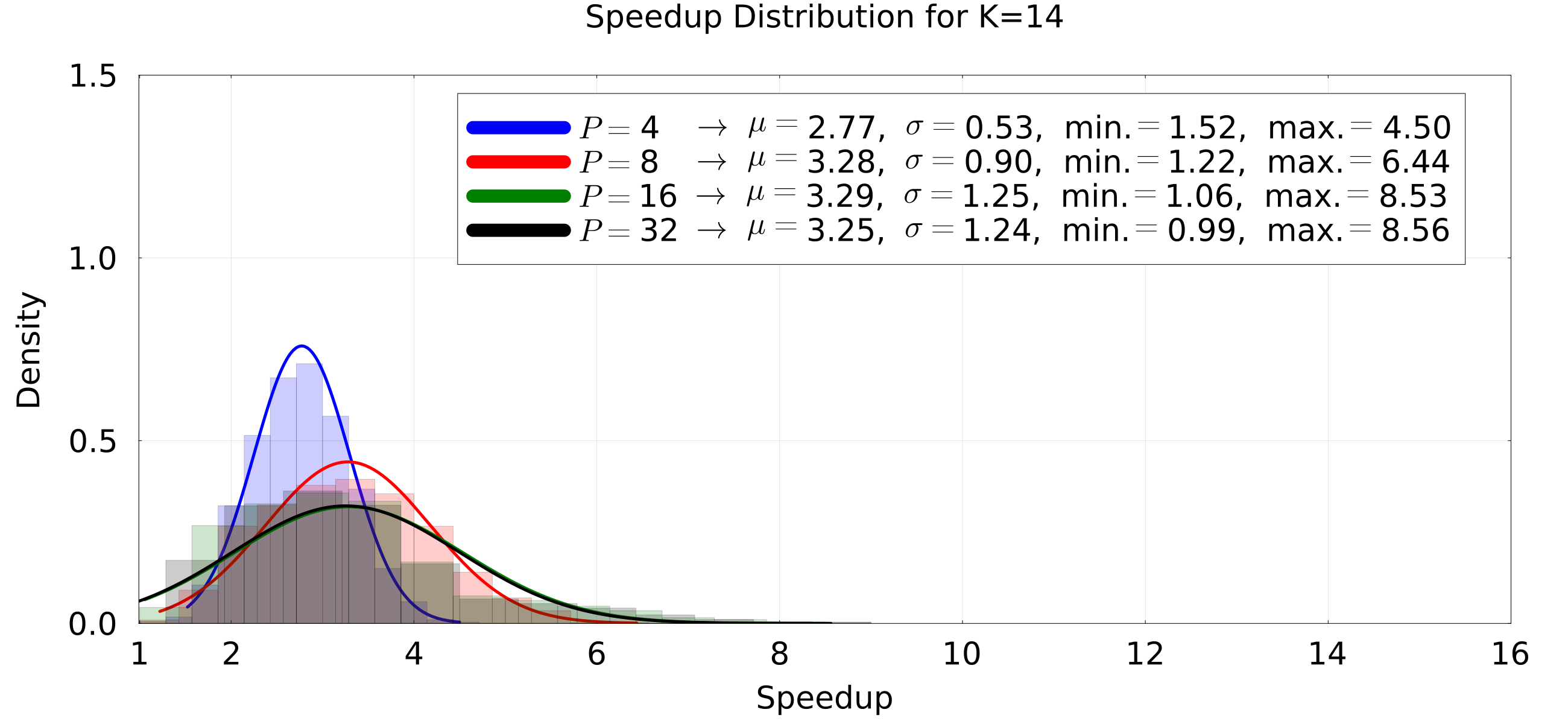}
        \label{fig:sub_e}
    \end{subfigure}
    \hfill 
    \begin{subfigure}[b]{0.49\textwidth}
        \centering
        \includegraphics[width=\textwidth]{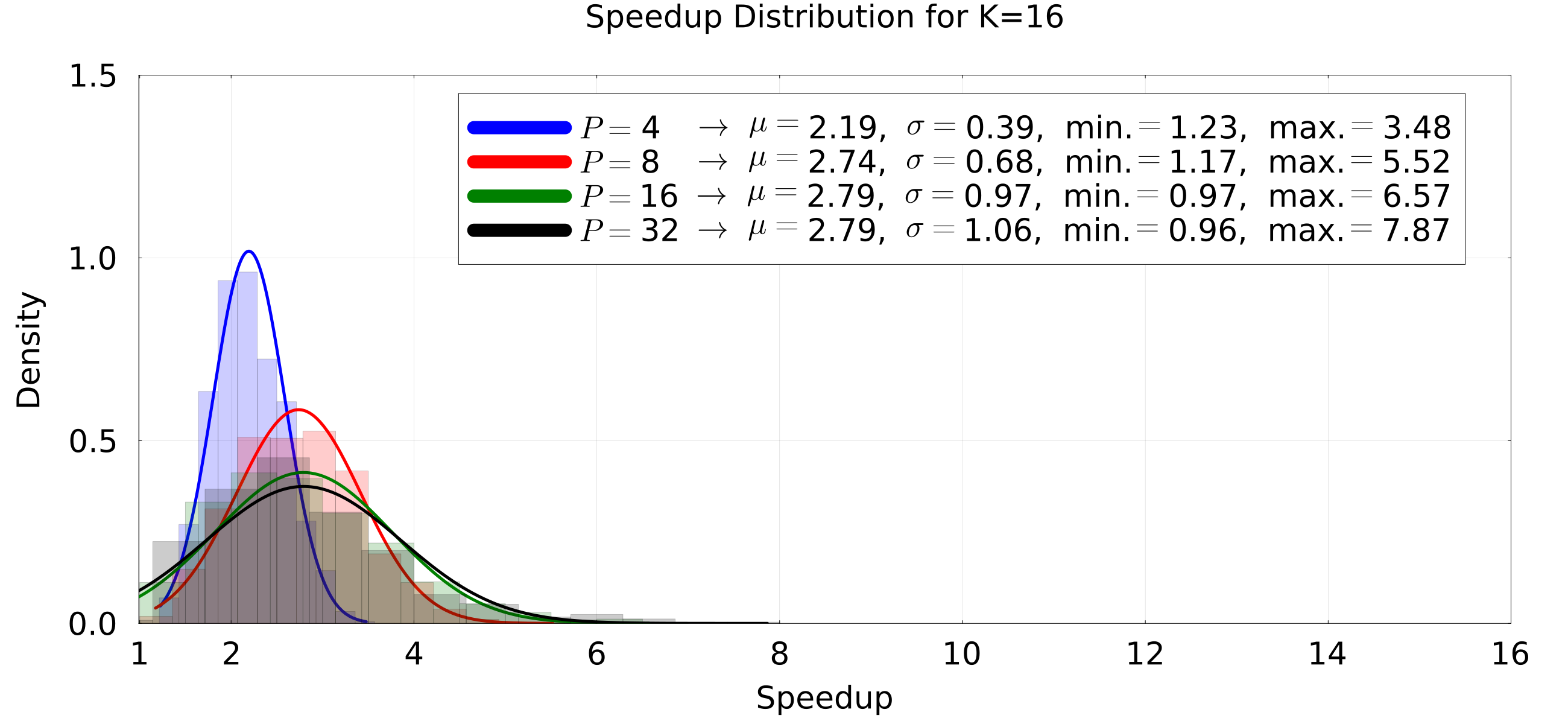}
        \label{fig:sub_f}
    \end{subfigure}
    
    \caption{\reviewertwo{Distribution of \reviewerone{speed-up} for \reviewerone{all $1000$} sample configurations depending on different values of $K$ and $P$. 
    }}
    \label{fig:results_sip}
\end{figure}

Considering one sample $j\in\{1,\ldots,1000\}$, let the \reviewerone{runtime}  be $C_j^{(S)}$ for the standard \reviewerone{\NewAnswers{AQS}} method and $C_j^{(P,K)}$ using $P\in\{4,8,16,32\}$ \reviewerone{and $K \in \{1,2,6,10,14,16\}$,} the \reviewerone{\NewAnswers{runtime}} for the parallel stepping method provided by Algorithm~\ref{alg:parallel}. We \reviewerone{then quantify} the \reviewerone{speed-up} as:
\begin{align}
    S_j^{(P,K)}=\frac{C_j^{(S)}}{C_j^{(P,K)}}.
\end{align}
\reviewerone{\NewAnswers{
Further, we introduce the average speed up for a fixed set of $P$ and $K$ as
\begin{align}
    \mu^{(P,K)} \; = \; \frac{1}{1000}\sum_{j=1}^{1000} \; S_j^{(P,K)}\;,
\end{align} 
and the standard deviation as
\begin{align}
    \sigma^{(P,K)}\;=\; \sqrt{\frac{1}{1000}\sum_{j=1}^{1000} \; \left(S_j^{(P,K)}-\mu^{(P,K)}\right)^2}\;.
\end{align}
}}
The results of the achieved \reviewerone{speed-up} is illustrated in Figure \ref{fig:results_sip}, where $S_{j}^{(P,K)}$ is represented in \reviewerone{\NewAnswers{histograms and a normal distribution according to the respective parameters $\mu^{(P,K)}$ and $\sigma^{(P,K)}$ is fitted for better visual interpretability.}}
In all cases, we observe a considerable advantage when using the \reviewerone{\NewAnswers{proposed parallel method}}.

\reviewerone{\NewAnswers{For any combination of $P$ and $K$ the lowest average speed up was observed as $\mu^{(4,1)} = 2.02$, which still cuts the average computational cost in half. For other combinations of $P$ and $K$ the speed up is higher with the maximum at $\mu^{(32,10)} = 6.33$.}} We emphasize that all speed-ups come without any loss in numerical or approximation accuracy.\\

\begin{figure}[htbp]
    \centering
    
    \begin{subfigure}[b]{0.49\textwidth}
        \centering
        \includegraphics[width=\textwidth]{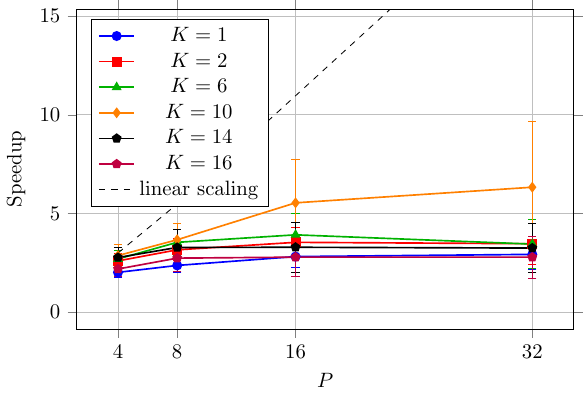}
        \caption{Speed-up regression on number of threads $P$}
        \label{fig:res_sub_a}
    \end{subfigure}
    \hfill 
    \begin{subfigure}[b]{0.49\textwidth}
        \centering
        \includegraphics[width=\textwidth]{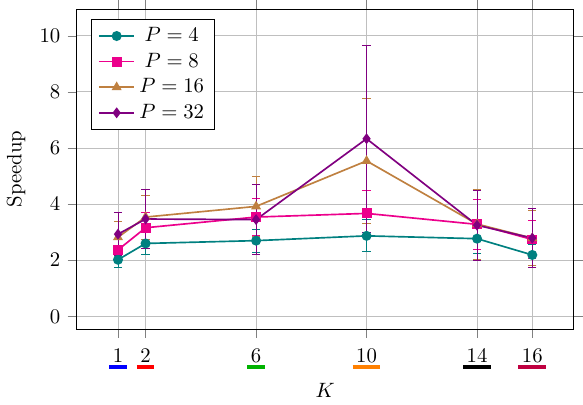}
        \caption{Speed-up regression on parameter $K$}
        \label{fig:res_sub_b}
    \end{subfigure}
    \caption{\reviewerone{Comparison of speed-up when looking at the main hyper-parameters $P$ and $K$.}}
    \label{fig:aqs_results}
\end{figure}

\reviewerone{\NewAnswers{
In Figure \ref{fig:res_sub_b} we can see that the optimal choice of $K$ lies somewhere between $K=6$ and $K=14$, while $K=10$ gives us the strongest result for the benchmark problem. This is in line with our general expectation and intuition of the parameter $K$. While small values of $K$ carry a large overhead considering the additional minimization starting from $\bm{r}_{iK}^c$ for any $i=1,\dots,P-1$, larger values of $K$ carry an increased risk of miss-predictions. 
The latter is caused by a large shear $KP\Delta\gamma$ relative to $\bm{r}_{\star}$ implying large affine shears at level \rom{I}, which, in turn, increases the chance that we end up in a different basin of attraction when compared to the standard AQS protocol.
\\
The optimal speed-up is only generated when a larger value of $K$ ($>2$) is combined with a considerable number of threads ($P=16,32$).
For the parameter $K$ we conclude that there is a sweet spot at $K=10$.
\\
When looking at the number of threads $P$ used in the parallel method we observe that the scaling is, as expected, not linear, which is shown by the results in Figure \ref{fig:res_sub_a}. Essentially, the same argument as for larger values of $K$ applies here too, the total relative shear of $KP\Delta\gamma$ becomes too large and the likelihood for miss-predictions of affine level \rom{I} steps increases. 
Additionally, one needs to keep in mind, that we ask of each thread to do the $K$ level \rom{II} steps $\left(\mathcal{F}\circ \bm{F}^{(\Delta\gamma )}\right)^K$. In a perfect world, each of the threads takes the same amount of time and comes back to the main thread simultaneously, as shown in Figure \ref{fig:aqs_sip2}. However, each operation $\mathcal{F}$ takes an a-priori unknown number of CG minimization steps, which is the main computational contributor to the runtime on an isolated thread. In consequence, threads stay idle until the thread taking the most amount of time, equivalently the one with largest number of total CG steps in the $K$ minimization procedures, terminates. 
This prevents effective load balancing in general. 
Yet, if one thread $i>0$ is providing a incorrect configuration all configuration provided by threads $j>i$ have to be rejected.
Therefore, part of this problem could be mitigated by establishing checking the validity of available incoming and outgoing configuration pairs live, and immediately stopping all threads $j>i$ if a miss-prediction on thread $i$ is detected.
However, this is hard to implement and does not completely  solve the load balancing problem. For this reason, we have not implemented this strategy. 
}}
\\

\subsection{Reproducibility}
All of the numerical work was done in the Julia programming language \cite{bezanson2017julia} and is available in the open source GitHub repository \textbf{Jumol} \cite{jumol_git}. The code is open source, and we invite everyone interested to contribute.

\section{Conclusion and future work}
In this paper, we presented a parallel stepping scheme offering a significant numerical speed-up without any loss in solution accuracy.\\
Our method builds on the concept that performing \reviewerone{\NewAnswers{parallel} AQS runs and then} individually connecting configurations at \reviewerone{\NewAnswers{the same shear value} resulting in a} parallelization of the deformation procedure \reviewerone{in the spirit of a predictor-corrector scheme}. 
\reviewerone{\NewAnswers{This procedure exploits that the potential energy landscape, most of the time changes moderately, with increasing shear}}. 
The \reviewerone{\NewAnswers{predictor-corrector}} process ensures that the landscape has not had any plastic or erratic changes, and the correct response trajectory is found.\\
We show that this method provides indeed significant speed-ups of at \reviewerone{\NewAnswers{least two-fold on average for the studied combinations of threads $P$ and $K$}} when compared to the conventional AQS method; however, this does not scale linearly with the number of threads $P$ as discussed and observed in Figure \ref{fig:aqs_results}. However, the aim of the algorithm is to consistently lower computational time when multiple threads are available\reviewerone{, and not to provide a perfectly scalable method}. 
This is done without any sacrifice in terms of accuracy compared to the same shear step size $\Delta\gamma$ of the standard AQS protocol. \reviewerone{As the overhead cost is negligible, the lower bound in terms of runtime of the proposed parallel method is the standard sequential AQS protocol.}\\
Our proposed method can be applied to a wide range of molecular systems to investigate inelastic deformation and fracture. Although we provided numerical benchmark examples investigating exclusively shear behavior of complex materials, the method can be easily extended to more general deformations as well as more complex deformation histories, such as cyclic deformation protocols.\\
Furthermore, the parallel stepping scheme provides a basis for a framework that efficiently finds multiple sets of alternative solution trajectories. Such alternative paths in the potential energy landscape are not relevant when performing athermal overdamped simulations but might be visited when performing molecular simulations at finite temperatures.
This way, further steps should be \reviewerone{taken} to develop numerically efficient strategies for finite temperature deformation, where the final material response trajectory is a result of probabilistic considerations taking into account a set of multiple trajectory candidates.\\ \\
In future research, it is desirable to find ways to a-priori prevent miss-predictions in order to reliably achieve a speed-up of close to the factor $P$. \reviewerone{Further, dynamical} step-sizing \reviewerone{and dynamical choice of the parameter $K$} would possibly \reviewerone{enable a user to exploit more threads efficiently.}
Future work will also involve developing extrapolation methods for the standard AQS protocol to minimize the number of steps in the minimization procedure. This can also be used on each of the parallel threads, thereby speeding up both the parallel stepping method and the standard AQS method.

\section*{CRediT authorship contribution statement}
\textbf{Maximilian Reihn}: Writing - review \& editing, Writing - original draft, Numerical work, Methodology; \textbf{Franz Bamer}: Writing - review \& editing, Writing - original
draft, Numerical Work; \textbf{Benjamin Stamm}: Review \& editing, Methodology.

\section*{Data availability}
No data was used for the research described in the article. The code used is publicly available.

\section*{Declaration of interests}
The authors declare that they have no known competing financial interests or personal relationships that could have appeared to
influence the work reported in this paper.

\section*{Acknowledgment}
The authors acknowledge financial support by the Deutsche Forschungsgemeinschaft (DFG) under project number 523939420.

\bibliographystyle{elsarticle-num}

\end{document}